\documentclass[journal]{IEEEtran}
\usepackage{amsmath,amssymb,amsfonts}
\usepackage{bm}
\usepackage{booktabs}
\usepackage{array}
\usepackage{multirow}
\usepackage{algorithm}
\usepackage{algpseudocode}
\usepackage{xcolor}
\usepackage{graphicx}

\usepackage{cite}

\title{Joint Antenna Geometry and Transmit Covariance Design for Near‑Field Multicast ISAC with Pinching Antenna Arrays}

\author{Hui Yang, Hao Feng, Ebrahim Bedeer, Ming Zeng, Mengyao Wang, Gaojian Huang and Mengyan Huang
    \thanks{H. Yang is with Hunan Institute of Science and Technology, Yueyang, China (email: achelyal@163.com).}    
    
    \thanks{H. Feng is with Hunan Institute of Engineering, and Donghua University as well as Laval University (email: 1219001@mail.dhu.edu.cn).}

    \thanks{E. Bedeer is with University of Saskatchewan, Saskatoon, SK, Canada (email: e.bedeer@usask.ca).}
    
    \thanks{M. Zeng is with Laval University, Quebec City, Canada (email: ming.zeng@gel.ulaval.ca).}

    \thanks{M. Wang, G. Huang and M. Huang are with Henan Polytechnic University, Jiaozuo, China (email: wangmengyao0615@gmail.com, g.huang@hpu.edu.cn, and huangmengyan@hpu.edu.cn).}

    }

\begin{document}
\maketitle	

\begin{abstract}
Pinching antenna systems (PASS) provide a flexible waveguide-based architecture for reconfiguring wireless propagation environments and creating geometry-dependent radiating apertures. This paper investigates a near-field multicast integrated sensing and communication (ISAC) system enabled by a lossy multi-waveguide PASS, where a base station transmits a common message to multiple communication users while simultaneously sensing one or multiple targets. The pinching antenna (PA) positions along the waveguides and the feed-domain transmit covariance matrix are jointly designed to improve sensing accuracy under multicast communication constraints. We first develop a near-field multicast ISAC signal model that accounts for waveguide attenuation, equal-radiated-power operation, geometry-dependent free-space propagation, and monostatic sensing. Then, we derive the  Fisher information matrix (FIM) for target parameter estimation and obtain a compact projected-Jacobian representation by exploiting the block-diagonal PA transfer structure. This representation reveals how the PA geometry and transmit covariance jointly affect the Cram\'er--Rao bound (CRB). Based on this structure, we further characterize the per-waveguide and cross-waveguide FIM contributions, the loss-aperture tradeoff, the identifiability condition, and the communication-sensing phase conflict. To minimize the CRB, we formulate a joint PA-position and transmit-covariance optimization problem subject to a multicast rate constraint, a feed-power budget, and PA deployment constraints. An alternating optimization algorithm is developed, where the covariance subproblem is solved as a semidefinite program and the PA-position subproblem is handled by waveguide-wise block coordinate descent. Numerical results show that the proposed joint design achieves substantially lower CRB than benchmark schemes with fixed PA positions or diagonal covariance matrices, demonstrating the benefits of jointly exploiting PA geometry and cross-waveguide covariance for near-field multicast ISAC.
\end{abstract}


{\color{black}
\section{Introduction}

\IEEEPARstart{P}{inching} antenna systems (PASS) have recently emerged as a promising architecture for realizing reconfigurable wireless apertures. In contrast to conventional arrays with permanently fixed antenna locations, PASS form localized radiating points, referred to as pinching antennas (PAs), along dielectric waveguides. By activating or repositioning these radiating points, the effective aperture can be adapted to the propagation environment, thereby shortening access distances, improving line-of-sight (LoS) connectivity, and providing additional spatial degrees of freedom, particularly at high carrier frequencies. Recent tutorials and surveys have summarized the underlying architectures, radiation mechanisms, channel models, pinching beamforming, channel acquisition, and resource-allocation strategies for PASS \cite{Liu25Tutorial,Xu26Tutorial,Liu26Survey,Illi26Survey}. Their potential for coverage enhancement, near-field transmission, resource allocation, and integrated sensing and communication (ISAC) has also been highlighted in recent overview articles \cite{Liu26Magazine,wijewardhana2025,Qin26MagazineISAC,Yang26Magazine,Zeng25Magazine}, while experimental and hardware-oriented studies have demonstrated the feasibility of dielectric-waveguide-based radiation \cite{kawai15New,fukuda20Experimental,suzuki22pinching,li24Dual,li25millimeter}. These developments have motivated applications of PASS to coverage enhancement \cite{FengHao_WCL26}, physical-layer security \cite{musri25PLS,badarneh25PLS}, wireless power transfer \cite{Gao26NOMAWPT,Li25}, non-orthogonal multiple access \cite{zeng25NOMARate,zeng2025EE,wang2024}, localization \cite{Feng_COMML26,he26SGISAC}, and ISAC.

Multicast transmission constitutes an important use case for PASS because a common data stream must be delivered reliably to a group of users, and the achievable multicast rate is fundamentally determined by the least favorable user \cite{Zhao_TCOM25,ren2023fundamental}. This worst-user limitation makes the spatial adaptability of PASS particularly attractive: the PA locations can potentially be configured to simultaneously improve several user channels rather than optimizing the link to a single receiver. Recent studies have therefore investigated PASS-assisted multicast transmission through PA placement, beamforming, waveguide operation, and blockage-aware deployment \cite{shan25Multicast,mu25pinching,Chen25Multicast,hanif26Blockage,Arnav26Multicast,shan25Group,shan26Multigroup}. These works, however, are primarily communication-oriented and do not address how the reconfigurable PASS geometry should be designed when the same transmitted waveform is also used for sensing.

In parallel, PASS have attracted considerable interest for sensing and ISAC because their reconfigurable radiating geometry directly controls the free-space propagation distances and phases observed by the sensing target. CRB-based sensing limits of PASS have been analyzed in \cite{ding25ISAC,Bozanis25ISAC,Jiang26Sensing}, while joint PA-position, transmit-power, and beamforming designs for different sensing--communication formulations have been studied in \cite{Qin25ISAC,zhang25ISAC,ouyang25ISAC,li25CRBISAC,li26ISAC,mao25ISAC}. PASS-assisted localization and channel-estimation methods have also exploited this geometry dependence \cite{he26SGISAC,liu25ISAC,Feng_COMML26,Pei26Localization}, and more recent works have considered segmented-waveguide architectures, low-altitude ISAC, and learning-based designs \cite{Guo25ISACUAV,Hu25ISACUAV,jiang25ISAC,gao26SegmentedISAC,JiangSegmentedISAC}. Nevertheless, most existing PASS-ISAC studies focus on unicast communication, employ restricted transmit structures, or optimize the antenna geometry only partially.

More closely related to the present work, PASS-aided integrated sensing and multicast communication has recently been studied in \cite{Shan26BCRB,Shan26ISAC}, where multicast performance and Bayesian sensing bounds are considered under different design criteria. These works establish the potential of PASS for jointly supporting group communication and sensing, but several fundamental issues remain insufficiently understood. First, in a multi-waveguide PASS, the transmit strategy is characterized not only by the power assigned to individual waveguides but also by the \emph{cross-waveguide correlations} represented by the off-diagonal entries of the feed-domain covariance matrix. How these correlations interact with the PA geometry in the sensing Fisher information matrix (FIM) has not been explicitly characterized. Second, the PA geometry affects both the guided-wave attenuation and the near-field amplitude/phase response. Consequently, enlarging the spatial aperture can improve geometric diversity but can simultaneously reduce the power delivered to distant PAs, creating a loss--aperture tradeoff that is absent in ideal lossless-array models. Third, the PA locations that favor coherent multicast transmission need not coincide with those providing a well-conditioned sensing FIM. These effects make independent geometry or power design inadequate for fully exploiting a lossy multi-waveguide PASS.

Motivated by these observations, we consider a near-field multicast ISAC system enabled by a lossy multi-waveguide PASS, where a base station (BS) transmits a common message to multiple communication users while simultaneously sensing one or multiple targets. The PA-position matrix $\mathbf{X}$ and the feed-domain transmit covariance matrix $\mathbf{Q}$ are jointly designed to minimize the sensing CRB subject to a multicast-rate requirement, a feed-power budget, and practical PA-deployment constraints. This problem differs fundamentally from conventional fixed-array ISAC because $\mathbf{X}$ simultaneously determines the free-space propagation geometry, the guided-wave phases, and the waveguide attenuation, whereas $\mathbf{Q}$ determines both the per-waveguide power allocation and the correlation among different waveguide signals. Hence, the two design variables are strongly coupled in both the communication channels and the sensing FIM. Rather than treating this coupling only as an optimization difficulty, we first expose its analytical structure and then use the resulting insights to guide the joint design.

The main contributions of this paper are summarized as follows.

\begin{itemize}

    \item \emph{Near-field multicast ISAC modeling with lossy multi-waveguide PASS:}
    We develop a unified signal model that incorporates reconfigurable PA geometry, guided-wave attenuation, equal-radiated-power operation, multicast downlink transmission, and monostatic near-field sensing. The multicast performance is characterized by the minimum achievable rate among all communication users, while the sensing accuracy is characterized through the CRB for the target locations and complex reflection coefficients. This formulation explicitly couples the PA positions and the feed-domain transmit covariance through both the communication and sensing models.

    \item \emph{Compact FIM representation and structural characterization:}
    We derive the sensing FIM and exploit the block-diagonal PA transfer structure to obtain a projected-Jacobian representation whose row dimension is reduced from $N_tN_r$ to $MN_r$. The resulting expression separates the dependence on the PA geometry from that on the feed-domain covariance and provides a convenient basis for both analysis and optimization. We further decompose the FIM into per-waveguide and cross-waveguide components, showing that the diagonal entries of $\mathbf{Q}$ determine the independent waveguide contributions, whereas its off-diagonal entries generate phase-sensitive cross-waveguide terms.

    \item \emph{PASS-specific sensing insights:}
    Based on the FIM structure, we characterize several mechanisms that are specific to the considered architecture. In particular, we show how intra-waveguide phase relationships can create constructive or destructive sensing contributions, identify the loss--aperture tradeoff produced by the competing effects of geometric diversity and guided-wave attenuation, establish the corresponding rank requirement for parameter identifiability, and reveal a communication--sensing phase conflict: PA locations favorable for coherent multicast signal accumulation do not, in general, provide the phase diversity preferred by near-field sensing.

    \item \emph{Joint geometry and full-covariance optimization:}
    We formulate the joint design of $\mathbf{X}$ and $\mathbf{Q}$ as a CRB-minimization problem subject to multicast-rate, total feed-power, and PA-placement constraints. For fixed PA positions, the FIM is affine in $\mathbf{Q}$, allowing the covariance subproblem to be reformulated as a convex semidefinite program via a Schur-complement representation. For fixed covariance, the nonconvex geometry subproblem is handled through waveguide-wise block coordinate descent with penalty-based feasibility control. These two steps are embedded into an alternating optimization framework.

    \item \emph{Numerical validation and design insights:}
    Numerical results demonstrate that jointly optimizing the PA geometry and the full covariance matrix consistently provides the lowest position-error bound among the considered schemes. The results distinguish the gains due to geometry optimization from the additional benefits of cross-waveguide covariance design, confirm the expected power-scaling and multicast communication--sensing tradeoff, and illustrate how waveguide length, propagation attenuation, and the numbers of waveguides and PAs affect the achievable sensing accuracy.

\end{itemize}

The remainder of this paper is organized as follows. Section~II presents the lossy multi-waveguide PASS model for multicast communication and near-field sensing. Section~III derives the sensing FIM and CRB and develops the structural insights underlying the proposed design. Section~IV formulates the joint PA-position and transmit-covariance optimization problem and presents the alternating optimization algorithm. Section~V provides numerical results, and Section~VI concludes the paper.
}

	\section{System Model}
	\label{sec:system_model}

	\subsection{PAs Geometry}
	
	We consider a PA-enabled ISAC multicast downlink in which a BS transmits a common message to $K$ single-antenna communication users (CUs) while illuminating $L$ sensing targets. The BS employs $M$ dielectric waveguides, each hosting $N$ PAs, giving a total of $N_{\mathrm{t}}=MN$ radiating elements. A co-located $N_{\mathrm{r}}$-element receive array enables monostatic sensing.
	
	The waveguides are deployed at height $h$ and are uniformly spaced along the $y$-axis with inter-waveguide spacing
	\begin{IEEEeqnarray}{rCl}
		d_y &=& \frac{D_y}{M-1},
		\label{eq:dy_def}
	\end{IEEEeqnarray}
	where $D_y$ is the total aperture of the waveguide array along the $y$-axis, i.e., the distance between the first and last waveguide. The $y$-coordinate of the feed point of the $m$-th waveguide is
	\begin{IEEEeqnarray}{rCl}
		y_m &=& (m-1)\,d_y, \qquad m=1,\ldots,M,
		\label{eq:ym_def}
	\end{IEEEeqnarray}
	so that the feed point is located at $\boldsymbol{\psi}_{m,0}=[0,\,y_m,\,h]^{\mathsf{T}}$. The $(m,n)$-th PA, i.e., the $n$-th PA of the $m$-th waveguide, occupies position
	\begin{IEEEeqnarray}{rCl}
		\boldsymbol{\alpha}_{m,n} &=& [x_{m,n},\; y_m,\; h]^{\mathsf{T}},
		\quad m=1,\ldots,M,\; n=1,\ldots,N, \IEEEeqnarraynumspace
		\label{eq:pa_location}
	\end{IEEEeqnarray}
	with $0<x_{m,n}\le D_x$, where $D_x$ is the waveguide length. The PA positions satisfy
	\begin{IEEEeqnarray}{rCl}
		0 &<& x_{m,1},\quad x_{m,N}\le D_x,
		\label{eq:ordering}
	\end{IEEEeqnarray}
	\begin{IEEEeqnarray}{rCl}
		x_{m,n}-x_{m,n-1} \ge \Delta_{\min}, \quad n=2,\ldots,N,
		\label{eq:min_sep}
	\end{IEEEeqnarray}
	where $\Delta_{\min}$ is the minimum separation between adjacent PAs on the same waveguide, required to suppress electromagnetic mutual coupling. We collect the $x$-coordinates of the PAs on waveguide $m$ in $\mathbf{x}_m=[x_{m,1},\ldots,x_{m,N}]^{\mathsf{T}}\in\mathbb{R}^{N}$, and stack all waveguides in the PAs position matrix
	\begin{IEEEeqnarray}{rCl}
		\mathbf{X} &=& [\mathbf{x}_1,\,\mathbf{x}_2,\,\ldots,\,\mathbf{x}_M]
		\in \mathbb{R}^{N\times M}.
		\label{eq:X_def}
	\end{IEEEeqnarray}
	
	The signal fed at the input of waveguide $m$ propagates along the dielectric medium before being radiated by each PA. This propagation is subject to attenuation. The power transmission factor between the $(n-1)$-th and $n$-th PA is \cite{shan25Group}
	\begin{IEEEeqnarray}{rCl}
		\kappa_{m,n}(\mathbf{x}_m)
		&=& 10^{-\varepsilon(x_{m,n}-x_{m,n-1})/10},
		\label{eq:kappa_def}
	\end{IEEEeqnarray}
	where $\varepsilon\ge 0$ (dB/m) is the waveguide attenuation coefficient. The power radiated by the $n$-th PA of waveguide $m$ is determined by the remaining waveguide power and a radiation coefficient $a_{m,n}\in(0,1]$ \cite{shan25Group}
	\begin{IEEEeqnarray}{rCl}
		P_{m,n}(\mathbf{x}_m)
		&=&
		a_{m,n}\; \kappa_{m,n}(\mathbf{x}_m) \Bigl(P_m - \sum_{j=1}^{n-1}P_{m,j}(\mathbf{x}_m)\Bigr),
		\label{eq:Pmn_def}
	\end{IEEEeqnarray}
	where $P_m$ is the feed power to waveguide $m$ and the total power radiated by waveguide $m$ is $P_{m}^{\rm tot}(\mathbf{x}_m)=\sum_{n=1}^{N}P_{m,n}(\mathbf{x}_m)\le P_m$. We adopt the equal power radiation model in which the radiation coefficients $a_{m,n}$ are chosen so that every PA on waveguide $m$ radiates the same power $P_{m}^{\rm tot}(\mathbf{x}_m)/N$. Hence, the required power radiation coefficients follow as \cite{shan25Group}
	\begin{IEEEeqnarray}{rCl}
		a_{m,n}
		&=& \frac{P_{m}^{\rm tot}(\mathbf{x}_m)}{%
			\kappa_{m,n}(\mathbf{x}_m)\bigl(N P_m -
			(n-1)\,P_{m}^{\rm tot}(\mathbf{x}_m)\bigr)}.
		\label{eq:amn_equal}
	\end{IEEEeqnarray}
	Please note that the equal radiated power $P_{m}^{\rm tot}(\mathbf{x}_m)/N$ value on waveguide $m$ depends on the PA positions $\mathbf{x}_m$ through the propagation loss factors $\kappa_{m,n}(\mathbf{x}_m)$. Also it is worthy to note that for the ideal case of a lossless waveguide, i.e., $\varepsilon=0$, the propagation factor reduces to $\kappa_{m,n}(\mathbf{x}_m)=1$ for all $m$ and $n$, meaning no feed power is dissipated in the waveguide. Consequently, all feed power is radiated, i.e., $P_{m}^{\rm tot}(\mathbf{x}_m)=P_m$.
	
	\subsection{Multicast Communication Model}
	
	At each channel use $t$, the BS transmits a multicast signal $\mathbf{u}(t)\sim\mathcal{CN}(\mathbf{0},\mathbf{Q}) \in\mathbb{C}^{M}$, where $\mathbf{Q}\succeq\mathbf{0}$ is the transmit signal covariance matrix whose $m$-th diagonal entry $[\mathbf{Q}]_{mm}=P_m$ represents the feed power allocated to waveguide $m$, subject to the total feed power budget $\operatorname{tr}(\mathbf{Q})=\sum_{m=1}^{M}P_m\le P$, where $P$ denotes the total power. The signal radiated by all $N_{\mathrm{t}}$ PAs is
	\begin{IEEEeqnarray}{rCl}
		\mathbf{x}(t) &=& \mathbf{G}(\mathbf{X})\,\mathbf{u}(t) \in\mathbb{C}^{N_{\mathrm{t}}},
		\label{eq:tx_signal}
	\end{IEEEeqnarray}
	where $\mathbf{G}(\mathbf{X})\in\mathbb{C}^{N_{\mathrm{t}}\times M}$ is the block-diagonal PA transfer matrix with $m$-th block
	\begin{IEEEeqnarray}{rCl}
		\mathbf{g}(\mathbf{x}_m)
		&=& \sqrt{\rho_m(\mathbf{x}_m)}
		\bigl[e^{-j\kappa_g x_{m,1}},\;\ldots,\;
		e^{-j\kappa_g x_{m,N}}\bigr]^{\mathsf{T}}
		\in\mathbb{C}^{N}, \nonumber \\ &&
		\label{eq:g_def}
	\end{IEEEeqnarray}
	where $\rho_m(\mathbf{x}_m) = {P_{m}^{\rm tot}(\mathbf{x}_m)}/{(NP_m)} \;\in\;(0,1/N]$ is the normalized loss factor, $\kappa_g=2\pi/\lambda_g$ is the guided wavenumber, and $\lambda_g=\lambda/n_{\mathrm{eff}}$. The full transfer matrix is
	\begin{IEEEeqnarray}{rCl}
		\mathbf{G}(\mathbf{X})
		&=&
		\operatorname{blkdiag}\!\bigl(
		\mathbf{g}(\mathbf{x}_1),\,\ldots,\,\mathbf{g}(\mathbf{x}_M)
		\bigr)
		\in \mathbb{C}^{N_{\mathrm{t}}\times M}, \IEEEeqnarraynumspace
		\label{eq:G_def}
	\end{IEEEeqnarray}
	and the radiated signal covariance is
	\begin{IEEEeqnarray}{rCl}
		\mathbf{S}_{\mathbf{x}}(\mathbf{X},\mathbf{Q})
		&=&
		\mathbb{E}\!\left\{\mathbf{x}(t)\mathbf{x}^{\mathsf{H}}(t)\right\}
		= \mathbf{G}(\mathbf{X})\,\mathbf{Q}\,\mathbf{G}^{\mathsf{H}}(\mathbf{X}) \in\mathbb{C}^{N_{\mathrm{t}}\times N_{\mathrm{t}}}. \nonumber \\ &&
		\label{eq:Sx_def}
	\end{IEEEeqnarray}
	
	Let $\boldsymbol{\psi}_k\in\mathbb{R}^{3}$ denote the location of CU $k$. The free-space LoS channel coefficient from the $(m,n)$-th PA to CU $k$ is
	\begin{IEEEeqnarray}{rCl}
		h_{m,n,k}(\mathbf{X})
		&=& \frac{\sqrt{\eta}}{d_{m,n,k}}\,
		e^{-j\kappa\, d_{m,n,k}},
		\label{eq:los_coeff}
	\end{IEEEeqnarray}
	where $d_{m,n,k}=\|\boldsymbol{\alpha}_{m,n}-\boldsymbol{\psi}_k\|_2$, $\kappa=2\pi/\lambda$, and $\eta=c^2/(16\pi^2 f_c^2)$. Define the channel vector from waveguide $m$ to CU $k$ as $\mathbf{h}_{m,k}(\mathbf{X})=[h_{m,1,k},\ldots,h_{m,N,k}]^{\mathsf{T}}\in\mathbb{C}^{N}$, and stack all waveguides to obtain the full channel coefficients vector from all the $N_{\rm t}$ PAs to CU $k$ as
	\begin{IEEEeqnarray}{rCl}
		\mathbf{h}_k(\mathbf{X})
		&=&
		\bigl[\mathbf{h}_{1,k}^{\mathsf{T}}(\mathbf{X}),\,\ldots,\,
		\mathbf{h}_{M,k}^{\mathsf{T}}(\mathbf{X})\bigr]^{\mathsf{T}}
		\in \mathbb{C}^{N_{\mathrm{t}}}.
		\label{eq:hk_def}
	\end{IEEEeqnarray}
	Then, the received signal at CU $k$ is
	\begin{IEEEeqnarray}{rCl}
		y_k(t) &=& \mathbf{h}_k^{\mathsf{T}}(\mathbf{X})\,\mathbf{x}(t) + z_k(t),
		\label{eq:yk}
	\end{IEEEeqnarray}
	where $z_k(t)\sim\mathcal{CN}(0,\sigma_k^2)$. The received signal power at CU $k$ is $\mathbb{E}\{|y_k(t)-z_k(t)|^2\} = \mathbf{h}_k^{\mathsf{T}}(\mathbf{X})\,\mathbf{S}_{\mathbf{x}}(\mathbf{X},\mathbf{Q})\,\mathbf{h}_k^{*}(\mathbf{X}) = \mathbf{h}_k^{\mathsf{H}}(\mathbf{X})\,\mathbf{S}_{\mathbf{x}}(\mathbf{X},\mathbf{Q})\,\mathbf{h}_k(\mathbf{X})$ for the Hermitian covariance $\mathbf{S}_{\mathbf{x}}\succeq\mathbf{0}$. Hence, under Gaussian signaling, the achievable rate of CU $k$ is \cite{ren2023fundamental}
	\begin{IEEEeqnarray}{rCl}
		R_k(\mathbf{X},\mathbf{Q})
		&=&
		\log_2\!\left(
		1+\frac{\mathbf{h}_k^{\mathsf{H}}(\mathbf{X})\,
			\mathbf{S}_{\mathbf{x}}(\mathbf{X},\mathbf{Q})\,
			\mathbf{h}_k(\mathbf{X})}
		{\sigma_k^2}
		\right),
		\label{eq:Rk}
	\end{IEEEeqnarray}
	and the multicast rate is the minimum achievable rate among all CUs:
	\begin{IEEEeqnarray}{rCl}
		R(\mathbf{X},\mathbf{Q})=\min_{k}R_k(\mathbf{X},\mathbf{Q}).
		\label{eq:rate_multicast}
	\end{IEEEeqnarray}
	As can be seen from \eqref{eq:rate_multicast}, the multicast communication performance is determined by both the PA geometry $\mathbf{X}$ and the multicast signal covariance $\mathbf{Q}$ through $\mathbf{S}_{\mathbf{x}}(\mathbf{X},\mathbf{Q})$.
	
	\subsection{Near-Field Sensing Model}
	
	The $L$ targets are assumed to lie in the near field of the PA system on the ground plane, and it is convenient that target $\ell$ is parameterized by its polar coordinates $(r_\ell,\theta_\ell)$ in the horizontal plane:
	\begin{IEEEeqnarray}{rCl}
		\mathbf{p}_\ell
		&=& \bigl[r_\ell\cos\theta_\ell,\; r_\ell\sin\theta_\ell\bigr]^{\mathsf{T}}
		\in\mathbb{R}^{2},\quad \ell=1,\ldots,L. \IEEEeqnarraynumspace
		\label{eq:target_pos}
	\end{IEEEeqnarray}
	We collect all target positions in $\mathbf{P}=[\mathbf{p}_1,\ldots,\mathbf{p}_L]\in\mathbb{R}^{2\times L}$. The corresponding near-field propagation distances to target $\ell$ are
	\begin{IEEEeqnarray}{rCl}
		d^{(\mathrm{t})}_{m,n,\ell}
		&=& \sqrt{\bigl(x_{m,n}-r_\ell\cos\theta_\ell\bigr)^2
			+\bigl(y_m-r_\ell\sin\theta_\ell\bigr)^2+h^2},\label{eq:nearfield_dists_tx} \IEEEeqnarraynumspace\\
		d^{(\mathrm{r})}_{q,\ell}
		&=& \sqrt{\bigl(x_{\mathrm{r},q}-r_\ell\cos\theta_\ell\bigr)^2
			+\bigl(y_{\mathrm{r},q}-r_\ell\sin\theta_\ell\bigr)^2+h_r^2},
		\label{eq:nearfield_dists_rx}
	\end{IEEEeqnarray}
	where $h_r$ is the height of the receive array. The transmit steering vector of target $\ell$, incorporating near-field amplitude and phase, is defined entry-wise as
	\begin{IEEEeqnarray}{rCl}
		\bigl[\mathbf{a}_{\mathrm{t}}(\mathbf{p}_\ell,\mathbf{X})\bigr]_{(m-1)N+n}
		&=& \frac{1}{d^{(\mathrm{t})}_{m,n,\ell}}\,
		e^{-j\kappa\, d^{(\mathrm{t})}_{m,n,\ell}},
		\label{eq:at_entry}
	\end{IEEEeqnarray}
	for $m=1,\ldots,M$, $n=1,\ldots,N$. The receive steering vector is defined entry-wise as
	\begin{IEEEeqnarray}{rCl}
		\bigl[\mathbf{a}_{\mathrm{r}}(\mathbf{p}_\ell)\bigr]_q
		&=& \frac{1}{d^{(\mathrm{r})}_{q,\ell}}\,
		e^{-j\kappa\, d^{(\mathrm{r})}_{q,\ell}},\quad q=1,\ldots,N_{\mathrm{r}}.
		\label{eq:ar_entry}
	\end{IEEEeqnarray}
	Let $\beta_\ell\in\mathbb{C}$ denote the effective complex response coefficient of target $\ell$, which captures its scattering strength and phase and is related to the target cross section \cite{ren2023fundamental}. The distance-dependent transmit- and receive-path losses are already incorporated into $\mathbf{a}_t(\mathbf{p}_\ell,\mathbf{X})$ and $\mathbf{a}_r(\mathbf{p}_\ell)$. 
    Then, the received echo at the $N_{\mathrm{r}}$-element receive array at snapshot $t$ is
	\begin{IEEEeqnarray}{rCl}
		\mathbf{y}_{\mathrm{r}}(t)
		&=& \sum_{\ell=1}^{L}\beta_\ell\,
		\mathbf{a}_{\mathrm{r}}(\mathbf{p}_\ell)\,
		\mathbf{a}_{\mathrm{t}}^{\mathsf{T}}(\mathbf{p}_\ell,\mathbf{X})\,
		\mathbf{x}(t)
		+\mathbf{z}_{\mathrm{r}}(t),
		\label{eq:echo}
	\end{IEEEeqnarray}
	where $\mathbf{z}_{\mathrm{r}}(t)\sim\mathcal{CN}(\mathbf{0},\sigma_r^2\mathbf{I}_{N_{\mathrm{r}}})$. Defining the receive and transmit array matrices from all $L$ targets as
	\begin{IEEEeqnarray}{rCl}
		\mathbf{A}_{\mathrm{r}}(\mathbf{P})
		&=&
		[\mathbf{a}_{\mathrm{r}}(\mathbf{p}_1),\ldots,
		\mathbf{a}_{\mathrm{r}}(\mathbf{p}_L)]
		\in\mathbb{C}^{N_{\mathrm{r}}\times L}, \label{eq:Ar_def}\\
		\mathbf{A}_{\mathrm{t}}(\mathbf{P},\mathbf{X})
		&=&
		[\mathbf{a}_{\mathrm{t}}(\mathbf{p}_1,\mathbf{X}),\ldots,
		\mathbf{a}_{\mathrm{t}}(\mathbf{p}_L,\mathbf{X})]
		\in\mathbb{C}^{N_{\mathrm{t}}\times L}, \label{eq:At_def}
	\end{IEEEeqnarray}
	and $\mathbf{B}=\operatorname{diag}(\beta_1,\ldots,\beta_L)$, the received sensing signal over all $T_{\mathrm{s}}$ snapshots is
	\begin{IEEEeqnarray}{rCl}
		\mathbf{Y}
		&=&
		\mathbf{A}_{\mathrm{r}}(\mathbf{P})\,\mathbf{B}\,
		\mathbf{A}_{\mathrm{t}}^{\mathsf{T}}(\mathbf{P},\mathbf{X})\,
		\mathbf{X}_{\mathrm{tx}}
		+\mathbf{Z},
		\label{eq:sensing_model}
	\end{IEEEeqnarray}
	where $\mathbf{Y}\in\mathbb{C}^{N_{\mathrm{r}}\times T_{\mathrm{s}}}$, $\mathbf{X}_{\mathrm{tx}}=[\mathbf{x}(1),\ldots,\mathbf{x}(T_{\mathrm{s}})]\in\mathbb{C}^{N_{\mathrm{t}}\times T_{\mathrm{s}}}$, and $\mathbf{Z}\in\mathbb{C}^{N_{\mathrm{r}}\times T_{\mathrm{s}}}$ has i.i.d.\ columns $\sim\mathcal{CN}(\mathbf{0},\sigma_r^2\mathbf{I})$. As shown later, under an average power constraint the dependence on $\mathbf{X}_{\mathrm{tx}}$ will be expressed in terms of the multicast signal covariance $\mathbf{Q}$ through $\mathbf{S}_{\mathbf{x}}(\mathbf{X},\mathbf{Q})$.

	\section{Fisher Information Matrix: Derivation and Insights}
	\label{sec:FIM}

	We assume that the BS knows the number of targets $L$ and seeks to estimate their near-field locations and complex reflection coefficients. The real-valued parameter vector to be estimated is
	\begin{IEEEeqnarray}{rCl}
		\boldsymbol{\xi}
		&=&
		\bigl[r_1,\theta_1,\ldots,r_L,\theta_L,\,
		\Re\{\beta_1\},\ldots,\Re\{\beta_L\}, \nonumber \\ && \hfill
		\Im\{\beta_1\},\ldots,\Im\{\beta_L\}
		\bigr]^{\mathsf{T}}
		\in\mathbb{R}^{4L}.
		\label{eq:xi_def}
	\end{IEEEeqnarray}

	\subsection{FIM Derivation}
	\label{subsec:fim_deriv}

	With the help of the vectorization--Kronecker identity, i.e., $\operatorname{vec}(\mathbf{M}\mathbf{N}) = (\mathbf{N}^{\mathsf{T}}\otimes \mathbf{I})\operatorname{vec}(\mathbf{M}),$ we can re-write the sensing model in \eqref{eq:sensing_model} as
	\begin{IEEEeqnarray}{rCl}
		\mathbf{y}
		&=&
		\bigl(\mathbf{X}_{\mathrm{tx}}^{\mathsf{T}}\otimes\mathbf{I}_{N_{\mathrm{r}}}\bigr)\,
		\bar{\boldsymbol{\mu}}(\boldsymbol{\xi},\mathbf{X})
		+\mathbf{z},
		\label{eq:y_vec}
	\end{IEEEeqnarray}
	where $\mathbf{y}=\operatorname{vec}(\mathbf{Y}) \in\mathbb{C}^{N_{\mathrm{r}}T_{\mathrm{s}}}$, $\mathbf{z}=\operatorname{vec}(\mathbf{Z})\sim\mathcal{CN}(\mathbf{0},\sigma_r^2\mathbf{I}_{N_{\mathrm{r}}T_{\mathrm{s}}})$, and
	\begin{IEEEeqnarray}{rCl}
		\bar{\boldsymbol{\mu}}(\boldsymbol{\xi},\mathbf{X})
		&=&
		\operatorname{vec}\!\bigl(
		\mathbf{A}_{\mathrm{r}}\mathbf{B}\mathbf{A}_{\mathrm{t}}^{\mathsf{T}}
		\bigr) \nonumber \\
		&=& \sum_{\ell=1}^{L}\beta_\ell\,
		\mathbf{a}_{\mathrm{t}}(\mathbf{p}_\ell,\mathbf{X})
		\otimes
		\mathbf{a}_{\mathrm{r}}(\mathbf{p}_\ell)
		\in\mathbb{C}^{N_{\mathrm{r}}N_{\mathrm{t}}}.
		\label{eq:mubar_sum}
	\end{IEEEeqnarray}
	Next, we define the Jacobian matrix as
	\begin{IEEEeqnarray}{rCl}
		\mathbf{D}_{\boldsymbol{\xi}}(\mathbf{X})
		&=&
		\frac{\partial\bar{\boldsymbol{\mu}}(\boldsymbol{\xi},\mathbf{X})}
		{\partial\boldsymbol{\xi}^{\mathsf{T}}}
		\in\mathbb{C}^{N_{\mathrm{r}}N_{\mathrm{t}}\times 4L},
		\label{eq:Dxi_def}
	\end{IEEEeqnarray}
	where the partial derivatives of $\bar{\boldsymbol{\mu}}(\boldsymbol{\xi},\mathbf{X})$ with respect to each entry of $\boldsymbol{\xi}$ are given in closed form in Appendix~\ref{app:jacobian}. Stacking the resulting columns in the same per-target order of \eqref{eq:xi_def} gives
	\begin{IEEEeqnarray}{rCl}
		\mathbf{D}_{\boldsymbol{\xi}}(\mathbf{X})
		&=&
		\Bigl[
		\underbrace{\mathbf{d}_{r_1},\,\mathbf{d}_{\theta_1}}_{\ell=1},\;
		\ldots,\;
		\underbrace{\mathbf{d}_{r_L},\,\mathbf{d}_{\theta_L}}_{\ell=L},\;
		\underbrace{\mathbf{A}_{\mathrm{t}}\diamond\mathbf{A}_{\mathrm{r}}}
		_{\Re\{\beta_\ell\}, \forall \ell},\;
		\underbrace{j\,\mathbf{A}_{\mathrm{t}}\diamond\mathbf{A}_{\mathrm{r}}}
		_{\Im\{\beta_\ell\}, \forall \ell}
		\Bigr], \nonumber \\
		\label{eq:Dxi_closed}
	\end{IEEEeqnarray}
	where $\diamond$ denotes the Khatri--Rao (column-wise Kronecker) product.
	
	Following \cite[eq.~(15.52)]{kay1993fundamentals}, the exact FIM is written as
	\begin{IEEEeqnarray}{rCl}
		\mathbf{J}_{\boldsymbol{\xi}}(\mathbf{X},\mathbf{X}_{\mathrm{tx}})
		&=&
		\frac{2}{\sigma_r^2}
		\Re\!\left\{
		\mathbf{D}_{\boldsymbol{\xi}}^{\mathsf{H}}(\mathbf{X})\,
		\bigl(\mathbf{X}_{\mathrm{tx}}^{*}\mathbf{X}_{\mathrm{tx}}^{\mathsf{T}}
		\otimes\mathbf{I}_{N_{\mathrm{r}}}\bigr)\,
		\mathbf{D}_{\boldsymbol{\xi}}(\mathbf{X})
		\right\}. \IEEEeqnarraynumspace
		\label{eq:FIM_exact_proved}
	\end{IEEEeqnarray}
	For sufficiently large $T_{\mathrm{s}}$, the sample covariance can be approximated by its expectation, i.e.,  $(1/T_{\mathrm{s}})\mathbf{X}_{\mathrm{tx}}\mathbf{X}_{\mathrm{tx}}^{\mathsf{H}}\approx \mathbf{S}_{\mathbf{x}}(\mathbf{X},\mathbf{Q})$, and since $\mathbf{S}_{\mathbf{x}}$ is Hermitian, $(1/T_{\mathrm{s}})\mathbf{X}_{\mathrm{tx}}^{*}\mathbf{X}_{\mathrm{tx}}^{\mathsf{T}}\approx \mathbf{S}_x^{\mathsf{T}}(\mathbf{X},\mathbf{Q})$. Hence, the average FIM  is given as
	\begin{IEEEeqnarray}{rCl}
		\mathbf{J}_{\boldsymbol{\xi}}(\mathbf{X},\mathbf{Q})
		&\approx&
		\frac{2T_{\mathrm{s}}}{\sigma_r^2}
		\Re\!\left\{
		\mathbf{D}_{\boldsymbol{\xi}}^{\mathsf{H}}(\mathbf{X})\,
		\bigl(\mathbf{S}_x^{\mathsf{T}}(\mathbf{X},\mathbf{Q})
		\otimes\mathbf{I}_{N_{\mathrm{r}}}\bigr)\,
		\mathbf{D}_{\boldsymbol{\xi}}(\mathbf{X})
		\right\}.\IEEEeqnarraynumspace
		\label{eq:FIM_Sx}
	\end{IEEEeqnarray}

	As can be seen in \eqref{eq:FIM_Sx}, to calculate $\mathbf{J}_{\boldsymbol{\xi}}(\mathbf{X},\mathbf{Q})$, we need to evaluate the Jacobian of dimension $N_{\mathrm{r}}N_{\mathrm{t}}\times 4L$ and the covariance matrix of the radiated signal of dimension $N_{\mathrm{t}}\times N_{\mathrm{t}}$, which involves high computational complexity. In the following, we exploit the block-diagonal structure of the PA transfer matrix $\mathbf{G}(\mathbf{X})$ to reduce the computational complexity of calculating the FIM $\mathbf{J}_{\boldsymbol{\xi}}(\mathbf{X},\mathbf{Q})$. This is achieved by substituting $\mathbf{S}_{\mathbf x}^{\mathsf{T}}(\mathbf{X},\mathbf{Q})=\mathbf{G}^{*}(\mathbf{X})\mathbf{Q}^{\mathsf{T}}\mathbf{G}^{\mathsf{T}}(\mathbf{X})$ into~\eqref{eq:FIM_Sx} and applying the mixed-product property $(\mathbf{AB})\otimes(\mathbf{CD})=(\mathbf{A}\otimes\mathbf{C})(\mathbf{B}\otimes\mathbf{D})$ to obtain
	\begin{IEEEeqnarray}{rCl}
		\mathbf{S}_x^{\mathsf{T}}\otimes\mathbf{I}_{N_{\mathrm{r}}}
		&=&
		\bigl(\mathbf{G}^{*}\otimes\mathbf{I}_{N_{\mathrm{r}}}\bigr)\,
		\bigl(\mathbf{Q}^{\mathsf{T}}\otimes\mathbf{I}_{N_{\mathrm{r}}}\bigr)\,
		\bigl(\mathbf{G}^{\mathsf{T}}\otimes\mathbf{I}_{N_{\mathrm{r}}}\bigr).
		\label{eq:kron_factored}
	\end{IEEEeqnarray}
	Then, by substituting \eqref{eq:kron_factored} into~\eqref{eq:FIM_Sx} and grouping the outer factors with $\mathbf{D}_{\boldsymbol{\xi}}$, the FIM becomes
	\begin{IEEEeqnarray}{rCl}
		\mathbf{J}_{\boldsymbol{\xi}}(\mathbf{X},\mathbf{Q})
		&=&
		\frac{2T_{\mathrm{s}}}{\sigma_r^2}
		\Re\!\left\{
		\widetilde{\mathbf{D}}_{\boldsymbol{\xi}}^{\mathsf{H}}(\mathbf{X})\,
		\bigl(\mathbf{Q}^{\mathsf{T}}\otimes\mathbf{I}_{N_{\mathrm{r}}}\bigr)\,
		\widetilde{\mathbf{D}}_{\boldsymbol{\xi}}(\mathbf{X})
		\right\},
		\label{eq:FIM_reduced}
	\end{IEEEeqnarray}
	where the \emph{projected Jacobian} $\widetilde{\mathbf{D}}_{\boldsymbol{\xi}}(\mathbf{X})$ is defined as
	\begin{IEEEeqnarray}{rCl}
		\widetilde{\mathbf{D}}_{\boldsymbol{\xi}}(\mathbf{X})
		&=&
		\bigl(\mathbf{G}^{\mathsf{T}}(\mathbf{X})\otimes\mathbf{I}_{N_{\mathrm{r}}}\bigr)\,
		\mathbf{D}_{\boldsymbol{\xi}}(\mathbf{X})
		\in\mathbb{C}^{MN_{\mathrm{r}}\times 4L}.
		\label{eq:Dtilde_def}
	\end{IEEEeqnarray}
	As can be seen in \eqref{eq:Dtilde_def}, the projected Jacobian $\widetilde{\mathbf{D}}_{\boldsymbol{\xi}}(\mathbf{X})$ has only $MN_{\mathrm{r}}$ rows (one per waveguide-receive antenna pair), while the original Jacobian $\mathbf{D}_{\boldsymbol{\xi}}(\mathbf{X})$ has $N_{\mathrm{t}}N_{\mathrm{r}}$ rows (one per transmit-receive antenna pair). Since $M \ll N_{\mathrm{t}}$, this projection represents a compression by a factor of $N$, which is only made possible by the block-diagonal structure of $\mathbf{G}(\mathbf{X})$. One can observe that the FIM in \eqref{eq:FIM_reduced} depends on PA position matrix $\mathbf{X}$ only through $\widetilde{\mathbf{D}}_{\boldsymbol{\xi}}(\mathbf{X})$, and on $\mathbf{Q}$ only through the term $\mathbf{Q}^{\mathsf{T}}\otimes\mathbf{I}_{N_{\mathrm{r}}}$.
	
    By the Cram\'{e}r--Rao theorem \cite{kay1993fundamentals}, the covariance matrix of any unbiased estimator $\hat{\boldsymbol{\xi}}$ of $\boldsymbol{\xi}$ 	satisfies
	\begin{IEEEeqnarray}{rCl}
		\mathbb{E}\!\left\{
		(\hat{\boldsymbol{\xi}}-\boldsymbol{\xi})
		(\hat{\boldsymbol{\xi}}-\boldsymbol{\xi})^{\mathsf{T}}
		\right\}
		&\succeq&
		\mathbf{J}_{\boldsymbol{\xi}}^{-1}(\mathbf{X},\mathbf{Q}),
		\label{eq:CRB_matrix}
	\end{IEEEeqnarray}
	i.e., the inverse FIM provides a lower bound on the estimation 	error covariance matrix. We adopt the trace of the inverse FIM 	as the sensing performance metric, which lower-bounds the total mean squared error (MSE) across all $4L$ estimated parameters:
	\begin{IEEEeqnarray}{rCl}
		\operatorname{CRB}(\mathbf{X},\mathbf{Q})
		&=&
		\operatorname{tr}\!\Bigl(
		\mathbf{J}_{\boldsymbol{\xi}}^{-1}(\mathbf{X},\mathbf{Q})
		\Bigr)
		\;\le\;
		\sum_{i=1}^{4L}
		\mathbb{E}\!\left\{(\hat{\xi}_i-\xi_i)^2\right\}. \IEEEeqnarraynumspace
		\label{eq:CRB_def}
	\end{IEEEeqnarray}
	Minimizing $\operatorname{CRB}(\mathbf{X},\mathbf{Q})$ (in the sense of a scalar performance metric) corresponds to minimizing a lower bound on the estimation error covariance, making it a common adopted sensing design criterion in the literature.
	
	After deriving the FIM and the CRB in \eqref{eq:FIM_reduced} and \eqref{eq:CRB_def}, respectively, our design goal is to jointly optimize the PA-position matrix $\mathbf{X}$ and the multicast signal covariance $\mathbf{Q}$ to minimize $\operatorname{CRB}(\mathbf{X},\mathbf{Q})$, subject to a minimum multicast rate $R(\mathbf{X},\mathbf{Q})\ge\bar{R}$ and a total feed power budget $\operatorname{tr}(\mathbf{Q})\le P$ constraints. Before proceeding to the solution, we exploit the structure of the FIM in \eqref{eq:FIM_reduced} to derive insights into the roles of $\mathbf{X}$ and $\mathbf{Q}$ in the CRB and how they affect the sensing performance and the communication-sensing tradeoff. These insights will be used to develop and reduce the complexity of the proposed algorithm.

	\subsection{Per-Waveguide and Cross-Waveguide Contributions to the FIM}
	\label{subsec:wg_decomp}
	
	In this subsection, we aim to quantify the per-waveguide and cross-waveguide contributions to the FIM in \eqref{eq:FIM_reduced}. We start by partitioning the projected Jacobian $\widetilde{\mathbf{D}}_{\boldsymbol{\xi}}(\mathbf{X})\in\mathbb{C}^{MN_{\mathrm{r}}\times 4L}$, defined in \eqref{eq:Dtilde_def}, into $M$ row-blocks of size $N_{\mathrm{r}}\times 4L$:
	\begin{IEEEeqnarray}{rCl}
		\widetilde{\mathbf{D}}_{\boldsymbol{\xi}}(\mathbf{X})
		&=&
		\bigl[
		\widetilde{\mathbf{D}}_1^{\mathsf{T}}(\mathbf{x}_1),\;
		\ldots,\;
		\widetilde{\mathbf{D}}_M^{\mathsf{T}}(\mathbf{x}_M)
		\bigr]^{\mathsf{T}}.
		\label{eq:Dtilde_partition}
	\end{IEEEeqnarray}
	To identify each block $\widetilde{\mathbf{D}}_m(\mathbf{x}_m)$ explicitly, we exploit the block-diagonal structure of $\mathbf{G}(\mathbf{X})$. Since $\mathbf{G}(\mathbf{X})= \operatorname{blkdiag}(\mathbf{g}(\mathbf{x}_1),\ldots,\mathbf{g}(\mathbf{x}_M))$, its transpose is also block-diagonal:
	\begin{IEEEeqnarray}{rCl}
		\mathbf{G}^{\mathsf{T}}(\mathbf{X})
		&=&
		\operatorname{blkdiag}\!\bigl(
		\mathbf{g}^{\mathsf{T}}(\mathbf{x}_1),\ldots,
		\mathbf{g}^{\mathsf{T}}(\mathbf{x}_M)
		\bigr),
		\label{eq:GT_blkdiag}
	\end{IEEEeqnarray}
	where $\mathbf{g}^{\mathsf{T}}(\mathbf{x}_m)\in\mathbb{C}^{1\times N}$ is the $m$-th row block. Applying the mixed-product property of the Kronecker product to the block-diagonal structure gives
	\begin{IEEEeqnarray}{rCl}
		\mathbf{G}^{\mathsf{T}}(\mathbf{X})\otimes\mathbf{I}_{N_{\mathrm{r}}}
		&=&
		\operatorname{blkdiag}\!\bigl(
		\mathbf{g}^{\mathsf{T}}(\mathbf{x}_1)\otimes\mathbf{I}_{N_{\mathrm{r}}},
		\ldots,
		\mathbf{g}^{\mathsf{T}}(\mathbf{x}_M)\otimes\mathbf{I}_{N_{\mathrm{r}}}
		\bigr),
		\nonumber\\
		\label{eq:GT_kron_blkdiag}
	\end{IEEEeqnarray}
	where each block $\mathbf{g}^{\mathsf{T}}(\mathbf{x}_m)\otimes\mathbf{I}_{N_{\mathrm{r}}}\in\mathbb{C}^{N_{\mathrm{r}}\times NN_{\mathrm{r}}}$.
	
	Similarly, the original Jacobian $\mathbf{D}_{\boldsymbol{\xi}}(\mathbf{X})\in\mathbb{C}^{N_{\mathrm{t}}N_{\mathrm{r}}\times 4L}$ can be re-defined in an $M$-block row-partition as
	\begin{IEEEeqnarray}{rCl}
		\mathbf{D}_{\boldsymbol{\xi}}(\mathbf{X})
		&=&
		\bigl[
		(\mathbf{D}_{\boldsymbol{\xi}}^{(1)})^{\mathsf{T}},\;
		\ldots,\;
		(\mathbf{D}_{\boldsymbol{\xi}}^{(M)})^{\mathsf{T}}
		\bigr]^{\mathsf{T}},
		\label{eq:D_partition}
	\end{IEEEeqnarray}
	where $\mathbf{D}_{\boldsymbol{\xi}}^{(m)}\in\mathbb{C}^{NN_{\mathrm{r}}\times 4L}$ is itself sub-partitioned into $N$ blocks of size $N_{\mathrm{r}}\times 4L$:
	\begin{IEEEeqnarray}{rCl}
		\mathbf{D}_{\boldsymbol{\xi}}^{(m)}
		&=&
		\bigl[
		(\mathbf{D}^{(m,1)}_{\boldsymbol{\xi}})^{\mathsf{T}},\;
		\ldots,\;
		(\mathbf{D}^{(m,N)}_{\boldsymbol{\xi}})^{\mathsf{T}}
		\bigr]^{\mathsf{T}},
		\label{eq:D_sub_partition}
	\end{IEEEeqnarray}
	where $\mathbf{D}^{(m,n)}_{\boldsymbol{\xi}}\in\mathbb{C}^{N_{\mathrm{r}}\times 4L}$ is the sub-block corresponding to the $(m,n)$-th PA. Substituting \eqref{eq:GT_kron_blkdiag} and \eqref{eq:D_partition} into \eqref{eq:Dtilde_def}, and with the help of the identity $(\mathbf{v}^{\mathsf{T}}\otimes\mathbf{I}_{N_{\mathrm{r}}})[\mathbf{A}_1^{\mathsf{T}},\ldots,\mathbf{A}_N^{\mathsf{T}}]^{\mathsf{T}}=\sum_{n}v_n\mathbf{A}_n$, the $m$-th block of $\widetilde{\mathbf{D}}_{\boldsymbol{\xi}}$ is given as
	\begin{IEEEeqnarray}{rCl}
		\widetilde{\mathbf{D}}_m(\mathbf{x}_m)
		&=&
		\bigl(
		\mathbf{g}^{\mathsf{T}}(\mathbf{x}_m)
		\otimes\mathbf{I}_{N_{\mathrm{r}}}
		\bigr)\,
		\mathbf{D}_{\boldsymbol{\xi}}^{(m)}
		\nonumber\\
		&=&
		\sqrt{\rho_m(\mathbf{x}_m)}
		\sum_{n=1}^{N}
		e^{-j\kappa_g x_{m,n}}\,
		\mathbf{D}^{(m,n)}_{\boldsymbol{\xi}}.
		\label{eq:Dtildem_def}
	\end{IEEEeqnarray}
	Please note that the sum in \eqref{eq:Dtildem_def} is a deterministic phase-weighted superposition, where the weights $\{e^{-j\kappa_g x_{m,n}}\}$ are fully determined by the PA positions $\{x_{m,n}\}$ and are therefore neither random nor unknown. However, deterministic phases do not automatically imply constructive reinforcement. Whether the $N$ per-PA Jacobian matrices $\{\mathbf{D}^{(m,n)}_{\boldsymbol{\xi}}\}$ reinforce or cancel each other in the sum depends on whether the entries of the per-PA Jacobian blocks combine constructively or destructively depending on their relative complex phases. That being said, the PA positions $\{x_{m,n}\}$ are therefore a design parameter to shape this superposition favorably.
	
	After constructing $\widetilde{\mathbf{D}}_{\boldsymbol{\xi}}(\mathbf{X})$ via \eqref{eq:Dtilde_partition} and \eqref{eq:Dtildem_def}, substituting into \eqref{eq:FIM_reduced} and noting that the $(m,m')$-th $N_{\mathrm{r}}\times N_{\mathrm{r}}$ block of $\mathbf{Q}^{\mathsf{T}}\otimes\mathbf{I}_{N_{\mathrm{r}}}$ equals $Q_{m'm}\mathbf{I}_{N_{\mathrm{r}}}$, the FIM decomposes as
	\begin{IEEEeqnarray}{rCl}
		\mathbf{J}_{\boldsymbol{\xi}}(\mathbf{X},\mathbf{Q})
		&=&
		\frac{2T_{\mathrm{s}}}{\sigma_r^2}
		\sum_{m=1}^{M}\sum_{m'=1}^{M}
		\Re\!\left\{
		Q_{m'm}\,
		\mathbf{\Psi}_{mm'}(\mathbf{X})
		\right\},
		\label{eq:FIM_decomp}
	\end{IEEEeqnarray}
	where $\mathbf{\Psi}_{mm'}(\mathbf{X})=\widetilde{\mathbf{D}}_m^{\mathsf{H}}(\mathbf{x}_m)\,\widetilde{\mathbf{D}}_{m'}(\mathbf{x}_{m'})\in\mathbb{C}^{4L\times 4L}$. By separating the diagonal and off-diagonal terms in \eqref{eq:FIM_decomp}, we obtain
	\begin{IEEEeqnarray}{rCl}
		\mathbf{J}_{\boldsymbol{\xi}}(\mathbf{X},\mathbf{Q})
		&=&
		\underbrace{
			\frac{2T_{\mathrm{s}}}{\sigma_r^2}
			\sum_{m=1}^{M}P_m\,
			\Re\!\bigl\{\mathbf{\Psi}_{mm}(\mathbf{X})\bigr\}
		}_{\text{depends only on the per-waveguide power }P_m}
		\nonumber\\
		&&+\;
		\underbrace{
			\frac{2T_{\mathrm{s}}}{\sigma_r^2}
			\sum_{m\neq m'}
			\Re\!\bigl\{
			Q_{m'm}\,\mathbf{\Psi}_{mm'}(\mathbf{X})
			\bigr\}
		}_{\text{depends on cross-waveguide correlation }Q_{m'm}}.
		\IEEEeqnarraynumspace
		\label{eq:FIM_incoherent_coherent}
	\end{IEEEeqnarray}
	Please note that the diagonal term in \eqref{eq:FIM_incoherent_coherent} depends only on the per-waveguide feed powers $P_m = [\mathbf{Q}]_{mm}$. It survives even when the signals fed to different waveguides are statistically independent, i.e., when $\mathbf{Q}$ is diagonal. Each waveguide contributes to the FIM independently, so no inter-waveguide signal correlation is required. On the other hand, the cross-waveguide term depends on the off-diagonal entries $Q_{m'm} = [\mathbf{Q}]_{m'm}$, $m \neq m'$, which represent the statistical correlation between the signals fed to waveguides $m$ and $m'$. A non-zero $Q_{m'm}$ means the two waveguide signals maintain a fixed phase relationship; whether this contributes constructively or destructively to the FIM depends on the phase of $Q_{m'm}$ relative to $\boldsymbol{\Psi}_{mm'}$. This provides a fundamental motivation for optimizing the full covariance matrix $\mathbf{Q}$, rather than only the per-waveguide powers $\{P_m\}$, to allow the cross-waveguide terms to contribute favorably.
	
	Substituting \eqref{eq:Dtildem_def} into $\mathbf{\Psi}_{mm}$ further expands the per-waveguide term as
	\begin{IEEEeqnarray}{rCl}
		\Re\{\mathbf{\Psi}_{mm}\}
		&=&
		\rho_m
		\sum_{n=1}^{N}
		\Re\!\left\{
		(\mathbf{D}^{(m,n)}_{\boldsymbol{\xi}})^{\mathsf{H}}
		\mathbf{D}^{(m,n)}_{\boldsymbol{\xi}}
		\right\}
		\nonumber\\
		&&+\;
		\rho_m
		\!\!\sum_{n=1}^{N} \sum_{\substack{n'=1\\ n'\neq n}}^{N}\!\!
		\Re\!\Big\{
		e^{-j\kappa_g(x_{m,n'}-x_{m,n})} \nonumber \\ && \hfill \times
		(\mathbf{D}^{(m,n)}_{\boldsymbol{\xi}})^{\mathsf{H}}
		\mathbf{D}^{(m,n')}_{\boldsymbol{\xi}}
		\Big\},
		\IEEEeqnarraynumspace
		\label{eq:Psi_mm_expanded}
	\end{IEEEeqnarray}
	and the cross-waveguide term as
	\begin{IEEEeqnarray}{rCl}
		\Re\{Q_{m'm}\mathbf{\Psi}_{mm'}\}
		&=&
		\sqrt{\rho_m\rho_{m'}}
		\sum_{n=1}^{N}\sum_{n'=1}^{N} \nonumber \\ && \hfill
		\Re\!\left\{
		Q_{m'm}\,
		e^{-j\kappa_g(x_{m',n'}-x_{m,n})}
		\right.
		\nonumber\\
		&&\quad\times\left.
		(\mathbf{D}^{(m,n)}_{\boldsymbol{\xi}})^{\mathsf{H}}
		\mathbf{D}^{(m',n')}_{\boldsymbol{\xi}}
		\right\}.
		\IEEEeqnarraynumspace
		\label{eq:Psi_mmp_expanded}
	\end{IEEEeqnarray}
	And one can observe the following from \eqref{eq:Psi_mm_expanded} and \eqref{eq:Psi_mmp_expanded}.

	\textit{1) PA self-information.} The first term in \eqref{eq:Psi_mm_expanded} ($n=n'$) accumulates the sensing information of each PA acting independently. Since each such term has the Gram matrix form $(\mathbf{D}^{(m,n)}_{\boldsymbol{\xi}})^{\mathsf{H}} 	\mathbf{D}^{(m,n)}_{\boldsymbol{\xi}} \succeq \mathbf{0}$, these terms are always positive semi-definite and cannot be cancelled by any choice of $\{x_{m,n}\}$.

    \textit{2) Inter-PA phase alignment:}
    The second term in~\eqref{eq:Psi_mm_expanded}, corresponding to $n\neq n'$, is modulated by the relative guided-wave phase $\kappa_g(x_{m,n'}-x_{m,n})$, which depends on the inter-PA spacings within waveguide $m$. These terms can contribute constructively or destructively depending on the alignment between this relative phase and the cross-information matrix $(\mathbf{D}_{\boldsymbol{\xi}}^{(m,n)})^{H} \mathbf{D}_{\boldsymbol{\xi}}^{(m,n')}$. Uniform spacing creates a structured set of relative phases whose diversity depends on $\kappa_g\delta$. When these phases sufficiently span the unit circle, the inter-PA terms can probe complementary sensing directions and improve the conditioning of $\boldsymbol{\Psi}_{mm}$. However, unfavorable spacings may produce repeated or nearly aligned phases, resulting in information redundancy. Therefore, both the PA aperture and the resulting relative phase pattern should be considered in the PA-position design.
	
	
	\textit{3) Cross-waveguide phase-alignment.} From \eqref{eq:Psi_mmp_expanded}, the cross-waveguide term is governed jointly by the geometric mean loss $\sqrt{\rho_m\rho_{m'}}$, the cross-waveguide phase $\kappa_g(x_{m',n'}-x_{m,n})$, and the phase of the inter-waveguide signal correlation $Q_{m'm}$. This reveals that the phase of $Q_{m'm}$ must be designed to compensate the total phase of $\mathbf{\Psi}_{mm'}$, which accumulates contributions from both the guided-wave path difference $\kappa_g(x_{m,n}-x_{m',n'})$ and the structure of the per-PA Jacobian blocks $\mathbf{D}^{(m,n)}_{\boldsymbol{\xi}}$. This shows that $\angle Q_{m'm}$ and the PA positions $\{x_{m,n}\}$, $\{x_{m',n'}\}$ must be \emph{jointly designed} to ensure constructive cross-waveguide contributions, providing a direct motivation for the joint $(\mathbf{Q},\mathbf{X})$ optimization of Section~\ref{sec:solution}.

	\subsection{Loss-Aperture Tradeoff}
	\label{subsubsec:loss_aperture}

    One can observe that the normalized loss factor $\rho_m(\mathbf{x}_m)$ globally attenuates the per-waveguide sensing contribution through every term in~\eqref{eq:Psi_mm_expanded}. Since $\rho_m$ depends on $\mathbf{x}_m$ only through the inter-PA spacings $\{\delta_{m,n}\}$ and decreases as the PAs are spread farther apart, increasing the aperture comes at the cost of higher propagation loss. Spreading the PAs farther apart enlarges the range of relative guided-wave phases $\{\kappa_g(x_{m,n'}-x_{m,n})\}_{n\neq n'}$, which can improve the conditioning of $\boldsymbol{\Psi}_{mm}$ when the resulting phases provide sufficiently diverse sensing directions. However, the improvement is not guaranteed, because certain spacings may produce repeated or unfavorably aligned phases. At the same time, increasing the PA aperture reduces $P_m^{\mathrm{tot}}(\mathbf{x}_m)$ through guided-wave propagation loss, thereby decreasing $\rho_m$. Conversely, clustering the PAs near the feed preserves the radiated-power factor but reduces the diversity of the relative guided-wave phases, potentially causing the inter-PA terms in~\eqref{eq:Psi_mm_expanded} to provide redundant sensing information. This loss--aperture tradeoff is a distinctive feature of the lossy PASS model and provides a key motivation for jointly optimizing the PA positions.
	
	
	To gain intuition on how loss limits the useful waveguide length, note that the power radiated by the $n$-th PA is bounded by $P_m^{\rm tot}(\mathbf{x}_m)/N \le P_m 10^{-\varepsilon x_{m,n}/10}/N$, where $x_{m,n} = \sum_{j=1}^{n}\delta_{m,j}$ is the cumulative distance from the feed. This power attenuation factor $10^{-\varepsilon x_{m,n}/10}$ decays exponentially with distance from the feed, meaning PAs placed far from the feed radiate negligible power and therefore contribute little to the FIM regardless of their positions. Specifically, the attenuation factor drops below $10^{-\alpha/10}$ of its feed value beyond a distance of $\alpha/\varepsilon$ meters, defining an effective useful waveguide length of $L_{\mathrm{eff}}=\alpha/\varepsilon$\,m. For example, setting $\alpha=10$\,dB gives $L_{\mathrm{eff}}=10/\varepsilon$\,m, beyond which PAs retain less than $10\%$ of their feed power. This provides a practical guideline for choosing the waveguide length $D_x$: setting $D_x\gg L_{\mathrm{eff}}$ wastes hardware without any meaningful sensing gain.

	\subsection{Parameter Identifiability}
	\label{subsubsec:identifiability}
	
	All $4L$ parameters in $\boldsymbol{\xi}$ are jointly identifiable if and only if $\mathbf{J}_{\boldsymbol{\xi}}(\mathbf{X},\mathbf{Q})\succ\mathbf{0}$. From \eqref{eq:FIM_reduced}, for any non-zero $\mathbf{v}\in\mathbb{R}^{4L}$:
	\begin{IEEEeqnarray}{rCl}
		\mathbf{v}^{\mathsf{T}}\mathbf{J}_{\boldsymbol{\xi}}\mathbf{v} &\propto& \bigl\|(\mathbf{Q}^{\mathsf{T}/2}\otimes\mathbf{I}_{N_{\mathrm{r}}}) \widetilde{\mathbf{D}}_{\boldsymbol{\xi}}\mathbf{v} \bigr\|^2. \label{eq:identifiability_quad}
	\end{IEEEeqnarray}
	This is strictly positive for all non-zero $\mathbf{v}$ if and only if $(\mathbf{Q}^{\mathsf{T}/2}\otimes\mathbf{I}_{N_{\mathrm{r}}})\widetilde{\mathbf{D}}_{\boldsymbol{\xi}}$ has full column rank equal to $4L$, which requires two conditions to hold simultaneously. First, $\mathbf{Q}\succ\mathbf{0}$: since $(\mathbf{Q}^{\mathsf{T}/2}\otimes\mathbf{I}_{N_{\mathrm{r}}})$ is invertible if and only if $\mathbf{Q}\succ\mathbf{0}$ by the mixed-product property of the Kronecker product. Second, $\operatorname{rank}(\widetilde{\mathbf{D}}_{\boldsymbol{\xi}}(\mathbf{X}))=4L$: given that $\mathbf{Q}\succ\mathbf{0}$ so that $(\mathbf{Q}^{\mathsf{T}/2}\otimes\mathbf{I}_{N_{\mathrm{r}}})$ is invertible and therefore rank-preserving, the expression $(\mathbf{Q}^{\mathsf{T}/2}\otimes\mathbf{I}_{N_{\mathrm{r}}})\widetilde{\mathbf{D}}_{\boldsymbol{\xi}}\mathbf{v}=\mathbf{0}$ if and only if $\widetilde{\mathbf{D}}_{\boldsymbol{\xi}}\mathbf{v}=\mathbf{0}$, so the full column rank of the product reduces to requiring $\operatorname{rank}(\widetilde{\mathbf{D}}_{\boldsymbol{\xi}}(\mathbf{X}))=4L$, which is where the PA geometry plays a critical role.
	
	Since $\widetilde{\mathbf{D}}_{\boldsymbol{\xi}}(\mathbf{X})\in\mathbb{C}^{MN_{\mathrm{r}}\times 4L}$, its rank satisfies $\operatorname{rank}(\widetilde{\mathbf{D}}_{\boldsymbol{\xi}}(\mathbf{X}))\le \min(MN_{\mathrm{r}}, 4L)$. For the identifiability condition $\operatorname{rank}(\widetilde{\mathbf{D}}_{\boldsymbol{\xi}}(\mathbf{X}))=4L$ to be achievable, we need $\min(MN_{\mathrm{r}}, 4L) \ge 4L$, which requires
	\begin{IEEEeqnarray}{rCl}
		MN_{\mathrm{r}} &\ge& 4L. \label{eq:dim_condition}
	\end{IEEEeqnarray}
    Note however that \eqref{eq:dim_condition} is only a necessary condition; the PA positions must still be chosen to ensure $\operatorname{rank}(\widetilde{\mathbf{D}}_{\boldsymbol{\xi}}(\mathbf{X}))=4L$, which is where the geometry design plays a critical role.

	\subsection{Communication--Sensing Phase Conflict}
	\label{subsubsec:comm_sensing_conflict}
	
	The PA positions $\{x_{m,n}\}$ affect both communication and sensing through the same phase weights $\{e^{-j\kappa_g x_{m,n}}\}$, yet the two objectives impose conflicting requirements on them.

	From~\eqref{eq:Rk} and~\eqref{eq:Sx_def}, the received signal power at CU~$k$ is
	$\mathbf{h}_k^{\mathsf{H}}\mathbf{S}_x(\mathbf{X},\mathbf{Q})
	\mathbf{h}_k = \mathbf{h}_k^{\mathsf{H}}\mathbf{G}(\mathbf{X})\mathbf{Q} 	\mathbf{G}^{\mathsf{H}}(\mathbf{X})\mathbf{h}_k$. Since $\mathbf{G}(\mathbf{X})$ is block-diagonal, substituting \eqref{eq:g_def} shows that the contribution of waveguide $m$ to
	this power is proportional to 
	\begin{IEEEeqnarray}{rCl}
		\rho_m P_m
		\Bigl|
		\sum_{n=1}^{N} e^{j\kappa_g x_{m,n}} h_{m,n,k}
		\Bigr|^2,
		\label{eq:comm_power_wg}
	\end{IEEEeqnarray}
	where the positive sign in the exponent arises from the conjugate in $\mathbf{g}^{\mathsf{H}}(\mathbf{x}_m)$. The expression in~\eqref{eq:comm_power_wg} is maximized when the per-PA contributions $e^{j\kappa_g x_{m,n}}h_{m,n,k}$ add
	\emph{coherently}. This demands a specific phase pattern dictated by the free-space
	channels $\{h_{m,n,k}\}$ and therefore tied to the geometry of
	CU~$k$.	In contrast, from~\eqref{eq:Psi_mm_expanded}, the sensing FIM benefits from relative phases $\{\kappa_g(x_{m,n'}-x_{m,n})\}_{n \neq n'}$ that span a wide range, so that the inter-PA cross terms probe different sensing subspaces and the conditioning of 	$\mathbf{\Psi}_{mm}$ is maximized, as discussed in Section~\ref{subsec:wg_decomp}.
	
	A phase pattern that achieves coherent combination for communication does not in general satisfy this spanning requirement for sensing, and vice versa. This fundamental point is one of the key motivation for the joint $(\mathbf{Q},\mathbf{X})$ optimization framework of 	Section~\ref{sec:solution}, where the two objectives are balanced through the multicast rate constraint~\eqref{eq:Ploc_rate}.

	\section{Proposed Joint Design}
	\label{sec:solution}

	In this section, we formulate and solve the joint PA-position and transmit-covariance design problem for CRB minimization under multicast-rate, feed-power, and PA deployment constraints.
	
	The block-diagonal structure of $\mathbf{G}(\mathbf{X})$ implies that its Gram matrix is diagonal: $\mathbf{\Lambda}(\mathbf{X}) = \mathbf{G}^{\mathsf{H}}(\mathbf{X})\mathbf{G}(\mathbf{X})= \operatorname{diag}(P_{1}^{\rm tot}/P_1,\ldots,P_{M}^{\rm tot}/P_M)$, with $0<P_{m}^{\rm tot}(\mathbf{x}_m)\le P_m$. Consequently, $\operatorname{tr}(\mathbf{S}_{\mathbf{x}})= \operatorname{tr}(\mathbf{Q}\mathbf{\Lambda}(\mathbf{X}))\le \operatorname{tr}(\mathbf{Q})\le P$, so the feed-power constraint $\operatorname{tr}(\mathbf{Q})\le P$ simultaneously bounds the total radiated power regardless of the PA positions. We jointly optimize the multicast signal covariance $\mathbf{Q}$ and the PA-position matrix $\mathbf{X}$ to minimize the CRB subject to a multicast rate floor and a power budget:
	\begin{IEEEeqnarray}{rCl}
		\mathcal{P}_{\mathrm{loc}}:\quad
		\min_{\mathbf{Q}\succeq\mathbf{0},\,\mathbf{X}}
		&& \operatorname{CRB}(\mathbf{X},\mathbf{Q}) \label{eq:Ploc_obj}\\
		\text{s.t.}\quad
		R(\mathbf{X},\mathbf{Q})&\ge&\bar{R},\label{eq:Ploc_rate}\\
		\operatorname{tr}(\mathbf{Q})&\le& P,\label{eq:Ploc_pow}\\
		0<x_{m,1},\;x_{m,N}&\le& D_x,\quad\forall m,\label{eq:Ploc_order}\\
		x_{m,n}-x_{m,n-1}&\ge&\Delta_{\min},\quad\forall m,\;n\ge 2.
		\label{eq:Ploc_sep}
	\end{IEEEeqnarray}
	For fixed $\mathbf{X}$, the projected Jacobian $\widetilde{\mathbf{D}}_{\boldsymbol{\xi}}(\mathbf{X})$ is a constant matrix, so the FIM in \eqref{eq:FIM_reduced} is \emph{linear} in $\mathbf{Q}$. This makes $\operatorname{CRB}(\mathbf{X},\mathbf{Q})=\operatorname{tr}(\mathbf{J}_{\boldsymbol{\xi}}^{-1})$ a convex function of $\mathbf{Q}$, and the subproblem over $\mathbf{Q}$ is a convex semidefinite programming (SDP) solvable to global optimality \cite{boyd2004convex}. For fixed $\mathbf{Q}$, the geometry update is handled by a waveguide-wise block coordinate descent (BCD) scheme. The two subproblems are solved in alternation.

	\subsection{Convex Covariance Subproblem for Fixed Geometry}
	\label{subsec:Q_subproblem}

	For fixed $\mathbf{X}$, we first recall the effective feed-domain channel of CU $k$,
	\begin{IEEEeqnarray}{rCl}
		\mathbf{z}_k(\mathbf{X})
		&=& \mathbf{G}^{\mathsf{H}}(\mathbf{X})\,\mathbf{h}_k(\mathbf{X})
		\in\mathbb{C}^{M},
		\label{eq:zk_def}
	\end{IEEEeqnarray}
	which allows the received signal power at CU $k$ to be written compactly as $\mathbf{h}_k^{\mathsf{H}}\mathbf{S}_x\mathbf{h}_k= \mathbf{z}_k^{\mathsf{H}}(\mathbf{X})\,\mathbf{Q}\,\mathbf{z}_k(\mathbf{X})$, so the achievable rate reduces to
	\begin{IEEEeqnarray}{rCl}
		R_k(\mathbf{X},\mathbf{Q})
		&=&
		\log_2\!\left(
		1+\frac{\mathbf{z}_k^{\mathsf{H}}(\mathbf{X})\,\mathbf{Q}\,
			\mathbf{z}_k(\mathbf{X})}
		{\sigma_k^2}
		\right).
		\label{eq:Rk_reduced}
	\end{IEEEeqnarray}
	The multicast rate constraint $R(\mathbf{X},\mathbf{Q})\ge\bar{R}$ is therefore equivalent to the set of linear constraints in $\mathbf{Q}$:
	\begin{IEEEeqnarray}{rCl}
		\operatorname{tr}\!\bigl(\mathbf{H}_k(\mathbf{X})\,\mathbf{Q}\bigr)
		&\ge& \gamma_{\mathrm{c},k},\quad k=1,\ldots,K,
		\label{eq:rate_trace}
	\end{IEEEeqnarray}
	where $\gamma_{\mathrm{c},k}=\sigma_k^2(2^{\bar{R}}-1)$ and $\mathbf{H}_k(\mathbf{X})=\mathbf{z}_k(\mathbf{X})\mathbf{z}_k^{\mathsf{H}}(\mathbf{X})$.
	
	We introduce an auxiliary variable $\mathbf{T}\succeq\mathbf{0}$ as an upper bound on $\mathbf{J}_{\boldsymbol{\xi}}^{-1}$ and apply the Schur complement lemma \cite{boyd2004convex} to reformulate $\mathbf{T}\succeq\mathbf{J}_{\boldsymbol{\xi}}^{-1}$ as the linear matrix inequality (LMI)
	\begin{IEEEeqnarray}{rCl}
		\begin{bmatrix}
			\mathbf{J}_{\boldsymbol{\xi}}(\mathbf{X},\mathbf{Q}) & \mathbf{I}_{4L}\\
			\mathbf{I}_{4L} & \mathbf{T}
		\end{bmatrix}
		&\succeq& \mathbf{0},
		\label{eq:schur_lmi}
	\end{IEEEeqnarray}
	which is jointly linear in $\mathbf{Q}$ and $\mathbf{T}$. The covariance subproblem is then
	\begin{IEEEeqnarray}{rCl}
		\mathcal{P}_{\mathbf{Q}}(\mathbf{X}):\quad
		\min_{\mathbf{Q}\succeq\mathbf{0},\,\mathbf{T}\succeq\mathbf{0}}
		&& \operatorname{tr}(\mathbf{T})
		\label{eq:PQ_obj}\\
		\text{s.t.}\quad
		\begin{bmatrix}
			\mathbf{J}_{\boldsymbol{\xi}}(\mathbf{X},\mathbf{Q}) & \mathbf{I}_{4L}\\
			\mathbf{I}_{4L} & \mathbf{T}
		\end{bmatrix}
		&\succeq& \mathbf{0},
		\label{eq:PQ_schur}\\
		\operatorname{tr}\!\bigl(\mathbf{H}_k(\mathbf{X})\,\mathbf{Q}\bigr)
		&\ge& \gamma_{\mathrm{c},k},
		\quad k=1,\ldots,K,
		\label{eq:PQ_rate}\\
		\operatorname{tr}(\mathbf{Q})
		&\le& P.
		\label{eq:PQ_pow}
	\end{IEEEeqnarray}
	All constraints are LMIs or linear in $(\mathbf{Q},\mathbf{T})$, hence $\mathcal{P}_{\mathbf{Q}}(\mathbf{X})$ is an SDP solvable to global optimality by standard interior-point methods \cite{boyd2004convex}.

	\subsection{Geometry Subproblem for Fixed Covariance}
	\label{subsec:X_subproblem}

	For fixed $\mathbf{Q}$, both the effective channel $\mathbf{z}_k(\mathbf{X})$ and the projected Jacobian $\widetilde{\mathbf{D}}_{\boldsymbol{\xi}}(\mathbf{X})$ depend nonlinearly on $\mathbf{X}$, making the geometry update nonconvex. We apply a waveguide-wise BCD scheme: each waveguide $m$ is updated in sequence while all other waveguides are held fixed.
	
	Let $\mathbf{X}_{-m} = [\mathbf{x}_1,\ldots,\mathbf{x}_{m-1},\mathbf{x}_{m+1},\ldots,\mathbf{x}_M]\in\mathbb{R}^{N\times(M-1)}$ denote the PA-position matrix of all waveguides except waveguide $m$.
	For fixed $\mathbf{X}_{-m}$, the $m$-th single-waveguide subproblem is
	\begin{IEEEeqnarray}{rCl}
		\mathcal{P}_m:\quad
		\min_{\mathbf{x}_m \in \mathcal{C}_m}
		&& \operatorname{CRB}(\mathbf{x}_m,\mathbf{X}_{-m},\mathbf{Q})
		\label{eq:Pm_obj}\\
		\text{s.t.}\quad
		\varphi_k(\mathbf{x}_m,\mathbf{X}_{-m},\mathbf{Q})
		&\ge& \gamma_{\mathrm{c},k},
		\quad k=1,\ldots,K,
		\label{eq:Pm_rate}
	\end{IEEEeqnarray}
	where $\varphi_k=\mathbf{z}_k^{\mathsf{H}}\mathbf{Q}\mathbf{z}_k$ is the received power at CU $k$ and $\mathcal{C}_m=\{\mathbf{x}_m:\;0<x_{m,1},\;x_{m,N}\le D_x,\;x_{m,n}-x_{m,n-1}\ge\Delta_{\min}\}$ is the convex feasible set.
	
	Before applying coordinate descent, we exploit the FIM structure discussed earlier to obtain a good initial point for $\mathcal{P}_m$, exploiting the following decoupling property. Under the equal-power radiation model, $\kappa_{m,n}=10^{-\varepsilon\delta_{m,n}/10}$ depends only on the spacing $\delta_{m,n}=x_{m,n}-x_{m,n-1}$. Since $P_{m,n}$ and $P_{m}^{\rm tot}$ are computed recursively from $\kappa_{m,n}$, they depend only on $\{\delta_{m,n}\}$, while the phase weights $e^{-j\kappa_g x_{m,n}}=e^{-j\kappa_g\sum_{j=1}^n\delta_{m,j}}$ depend on cumulative sums of the spacings. Consequently, the total radiated power $P_{m}^{\rm tot}(\mathbf{x}_m)$ depends on $\mathbf{x}_m$ only through the inter-PA spacings, while the phase weights depend on the absolute positions. This decoupling motivates a two-stage placement strategy.
	
	Stage~1: Uniform spacings
	\begin{IEEEeqnarray}{rCl}
		\delta_{m,n}^{\star}
		&=& \frac{D_x}{N},\quad n=1,\ldots,N,
		\label{eq:uniform_spacing}
	\end{IEEEeqnarray}
	maximize the aperture and yield identical per-section propagation attenuation along waveguide~$m$. Indeed, since the segment transmission factor satisfies 	$\kappa_{m,n}(\mathbf{x}_m)
	=10^{-\varepsilon(x_{m,n}-x_{m,n-1})/10}
	=10^{-\varepsilon \delta_{m,n}/10}$, uniform spacing makes $\kappa_{m,n}$ constant for all $n$, i.e., $\kappa_{m,n}=10^{-\varepsilon(D_x/N)/10}$. Equivalently, each inter-PA segment incurs the same loss of $\varepsilon D_x/N$~dB, so no single segment dominates the attenuation.
	Note however that the cumulative loss from the feed still increases with~$n$ because
	$x_{m,n}=\sum_{j=1}^{n}\delta_{m,j}=nD_x/N$, so the available guided power decays exponentially with distance even under uniform spacing.
	
	{Stage~2: Phase-coherence refinement.}
	With the Stage~1 placement as a starting point, a 1D search over a global shift $\Delta x_m$ is performed:
	\begin{IEEEeqnarray}{rCl}
		\Delta x_m^{\star}
		&=&
		\arg\max_{\Delta x_m\in[0,D_x/N]}
		\operatorname{tr}\!\bigl(
		\mathbf{\Psi}_{mm}(\mathbf{x}_m^{\star}
		+\Delta x_m\mathbf{1})
		\bigr),
		\label{eq:phase_shift_opt}
	\end{IEEEeqnarray}
	where $\mathbf{1}=[1,\ldots,1]^{\mathsf{T}}$. Shifting all PAs by $\Delta x_m$ preserves the inter-PA spacings $\{x_{m,n}-x_{m,n-1}\}$ for ${n\ge2}$, and therefore maintains the relative phase differences $\{\kappa_g(x_{m,n'}-x_{m,n})\}$ that govern inter-PA interference patterns. However, such a shift changes the absolute positions $\{x_{m,n}\}$, and in particular the distance from the feed to the first PA, $x_{m,1}$. As a result, the cumulative propagation loss along the waveguide and hence the total radiated power $P_m^{\rm tot}(\mathbf{x}_m)$ are generally affected by $\Delta x_m$.
	Therefore, Stage~2 introduces a controlled tradeoff: it refines the phase alignment of the per-PA contributions through absolute-position shifts, while potentially modifying the overall loss factor $\rho_m(\mathbf{x}_m)$. The search interval $[0,D_x/N]$ spans one segment of the waveguide phase and is solved efficiently via a one-dimensional bounded line search (e.g., {fminbnd} in Matlab).
	
	Then, the resulting initial positions are
	\begin{IEEEeqnarray}{rCl}
		x_{m,n}^{(0)}
		&=& (n-1)\,\frac{D_x}{N}+\Delta x_m^{\star},
		\quad n=1,\ldots,N,
		\label{eq:warm_start}
	\end{IEEEeqnarray}
	which spread PAs uniformly over the full waveguide length and align the phases, resulting in a significantly larger virtual aperture, faster convergence, and lower final CRB values.
	
	To handle the constraints of $\mathcal{P}_m$, we absorb them into the objective via an $\ell_1$ penalty:
	\begin{IEEEeqnarray}{rCl}
		\Phi_m(\mathbf{x}_m,\mu)
		&=&
		\operatorname{CRB}(\mathbf{x}_m,\mathbf{X}_{-m},\mathbf{Q}) \nonumber \\ &&
		+\mu\sum_{k=1}^{K}
		\bigl[\gamma_{\mathrm{c},k}
		-\varphi_k(\mathbf{x}_m,\mathbf{X}_{-m},\mathbf{Q})
		\bigr]_{+},
		\label{eq:Phi_exact}
	\end{IEEEeqnarray}
	where $[u]_{+} = \max\{0,u\}$ and $\mu>0$ is the penalty parameter. The reformulated problem is $\widetilde{\mathcal{P}}_m:\;\min_{\mathbf{x}_m\in\mathcal{C}_m}\Phi_m(\mathbf{x}_m,\mu)$. In practice, $\mu$ is increased until the maximum rate violation falls below a tolerance $\eta>0$. 
	Problem $\widetilde{\mathcal{P}}_m$ is solved by coordinate descent: at each inner iteration, coordinates $n=1,\ldots,N$ are visited in sequence and $x_{m,n}$ is updated by minimizing $\Phi_m$ over the interval $[\underline{x}_{m,n},\overline{x}_{m,n}]$, where
	\begin{IEEEeqnarray}{rCl}
		\underline{x}_{m,n}
		&=&
		\begin{cases}
			0, & n=1,\\
			x_{m,n-1}+\Delta_{\min}, & n\ge 2,
		\end{cases}
		\label{eq:lb_xmn}\\
		\overline{x}_{m,n}
		&=&
		\begin{cases}
			x_{m,n+1}-\Delta_{\min}, & n\le N-1,\\
			D_x, & n=N.
		\end{cases}
		\label{eq:ub_xmn}
	\end{IEEEeqnarray}
	The coordinate update is
	\begin{IEEEeqnarray}{rCl}
		x_{m,n}^{(s+1)}
		&\leftarrow&
		\arg\min_{x\in[\underline{x}_{m,n},\overline{x}_{m,n}]}
		\phi_{m,n}(x),
		\label{eq:coord_update}
	\end{IEEEeqnarray}
	where $\phi_{m,n}(x)$ is the restriction of $\Phi_m$ to the $n$-th coordinate. The function $\phi_{m,n}$ is continuous and piecewise smooth on the bounded interval and is solved a one-dimensional bounded line search (e.g., fminbnd in Matlab). Each coordinate update refreshes only the entries of $\widetilde{\mathbf{D}}_{\boldsymbol{\xi}}$ and $\mathbf{z}_k$ corresponding to the $n$-th PA of waveguide $m$; all other quantities remain unchanged. Since $\phi_{m,n}(x)$ may be multimodal, a one-dimensional search converges to a local minimizer. To mitigate local minima, the interval can be partitioned into subintervals with the best solution retained, or a multi-start strategy applied.

	\subsection{Overall Algorithm}
	\label{subsec:algorithm}

	The initial covariance is set to uniform isotropic allocation: $\mathbf{Q}^{(0)} = (P/M)\mathbf{I}_M$. The initial PA positions are provided by \eqref{eq:warm_start}. The overall procedure is summarized in Algorithm~\ref{alg:overall}.
	
	\begin{algorithm}[t]
		\caption{Alternating optimization for PASS ISAC localization design}
		\label{alg:overall}
		\begin{algorithmic}[1]
			\Require $\bar{R}$, $P$, $\Delta_{\min}$, $D_x$, penalty growth factor $\tau>1$, feasibility tolerance $\eta>0$, convergence thresholds $\epsilon_{\mathrm{out}}$, $\epsilon_{\mathrm{in}}$
			\State $\gamma_{\mathrm{c},k}\leftarrow \sigma_k^2(2^{\bar{R}}-1)$
			\State Compute $\Delta x_m^{\star}$ via \eqref{eq:phase_shift_opt} for each $m$; set $\mathbf{X}^{(0)}$ via \eqref{eq:warm_start}
			\State Set $\mathbf{Q}^{(0)}\leftarrow(P/M)\mathbf{I}_M$, $\mu\leftarrow\mu^{(0)}$; $t\leftarrow 0$
			\Repeat
			\State \textbf{Q-update:} Solve SDP $\mathcal{P}_{\mathbf{Q}}(\mathbf{X}^{(t)})$ \eqref{eq:PQ_obj}--\eqref{eq:PQ_pow}; set $\mathbf{Q}^{(t+1)}\leftarrow$ optimal $\mathbf{Q}$
			\State \textbf{X-update:}
			\For{$m=1$ \textbf{to} $M$}
			\State $\mathbf{x}_m\leftarrow\mathbf{x}_m^{(t)}$
			\Repeat \Comment{penalty loop}
			\Repeat \Comment{coordinate descent}
			\For{$n=1$ \textbf{to} $N$}
			\State Compute $[\underline{x}_{m,n},\overline{x}_{m,n}]$ via \eqref{eq:lb_xmn}--\eqref{eq:ub_xmn}
			\State $x_{m,n}\leftarrow\arg\min_{x\in[\underline{x}_{m,n},\overline{x}_{m,n}]}\phi_{m,n}(x)$ 
			\EndFor
			\Until{$\|\mathbf{x}_m-\mathbf{x}_m^{\mathrm{prev}}\|_2\le\epsilon_{\mathrm{in}}$}
			\If{$\max_k[\gamma_{\mathrm{c},k}-\varphi_k(\mathbf{x}_m;\mathbf{X}_{-m},\mathbf{Q}^{(t+1)})]_{+}\le\eta$}
			\State \textbf{break}
			\Else
			\State $\mu\leftarrow\tau\mu$, $\tau > 1$
			\EndIf
			\Until{feasibility met}
			\State $\mathbf{x}_m^{(t+1)}\leftarrow\mathbf{x}_m$
			\EndFor
			\State $t\leftarrow t+1$
			\Until{$|\operatorname{CRB}^{(t)}-\operatorname{CRB}^{(t-1)}|\le\epsilon_{\mathrm{out}}$}
			\Ensure $(\mathbf{X}^{(t)},\mathbf{Q}^{(t)})$
		\end{algorithmic}
	\end{algorithm}

\begin{figure}[t]
\centering
{\includegraphics[width=0.5\textwidth]{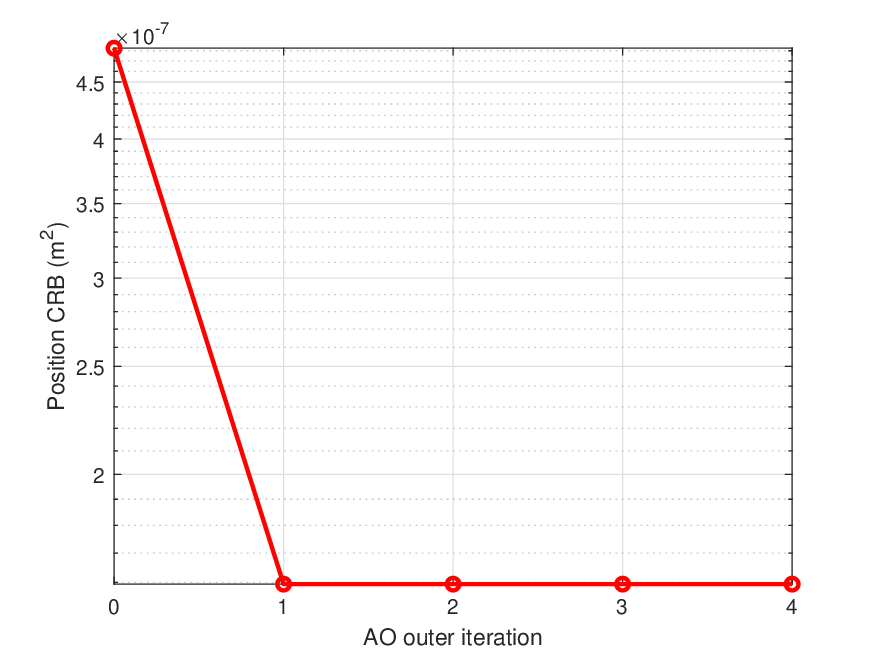}}
\caption{Convergence of the proposed alternating optimization (AO) algorithm at $P=30$ dBm. Iteration 0 denotes the geometry-informed feasible initialization, while each subsequent point is recorded after one complete covariance-and-position update.}
\label{fig:convergence}
\end{figure}

\begin{figure}[t]
\centering
{\includegraphics[width=0.5\textwidth]{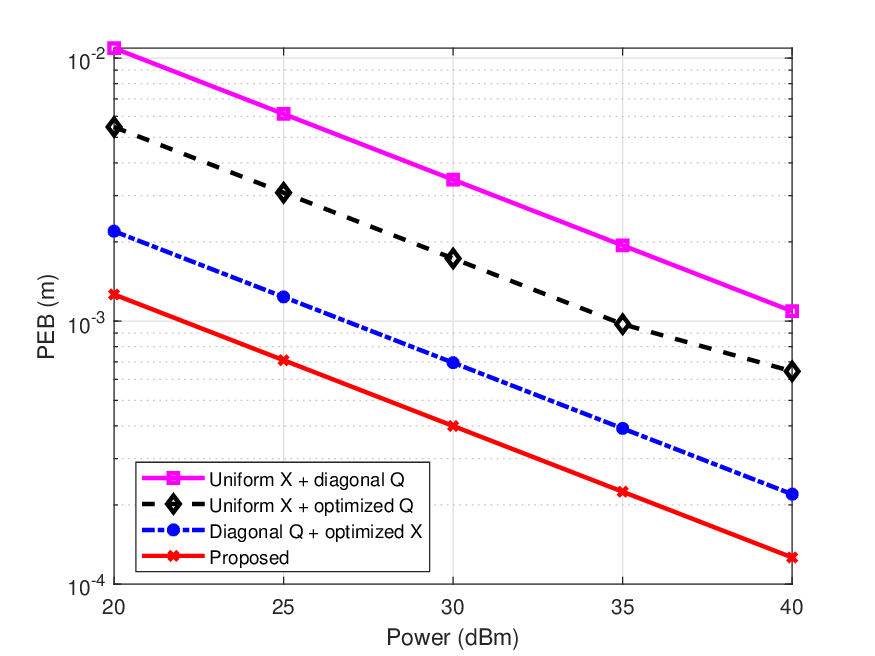}}
\caption{PEB versus the maximum transmit power, with $\bar R=1$ bit/s/Hz, $D_x=50$ m, and $\varepsilon=0.01$ dB/m.}
\label{fig:Power}
\end{figure}

\begin{figure}[t]
\centering
{\includegraphics[width=0.5\textwidth]{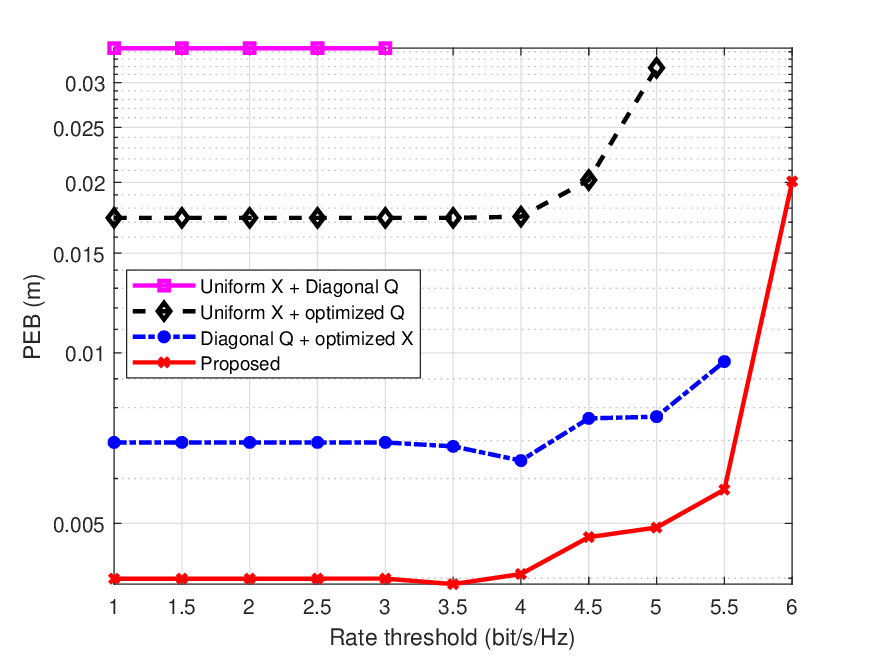}}
\caption{PEB versus the multicast-rate threshold, where $P=10$ dBm is used to highlight the communication--sensing tradeoff. Missing points indicate infeasible multicast-rate requirements.}
\label{fig:Rate}
\end{figure}

\begin{figure}[t]
\centering
{\includegraphics[width=0.5\textwidth]{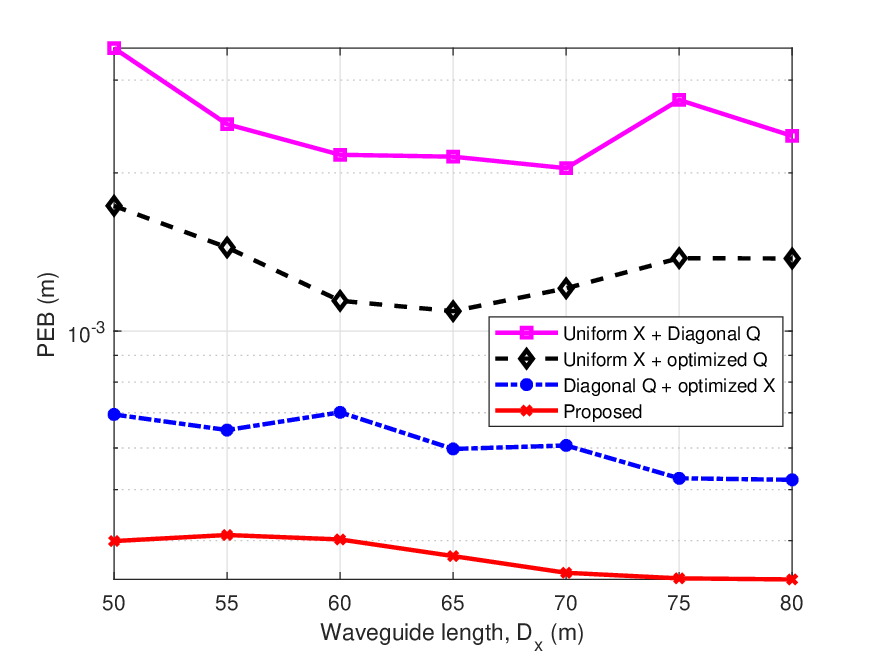}}
\caption{PEB versus the waveguide length, with $P=30$ dBm, $\bar R=1$ bit/s/Hz, and $\varepsilon=0.01$ dB/m.}
\label{fig:length}
\end{figure}

\begin{figure}[t]
\centering
{\includegraphics[width=0.5\textwidth]{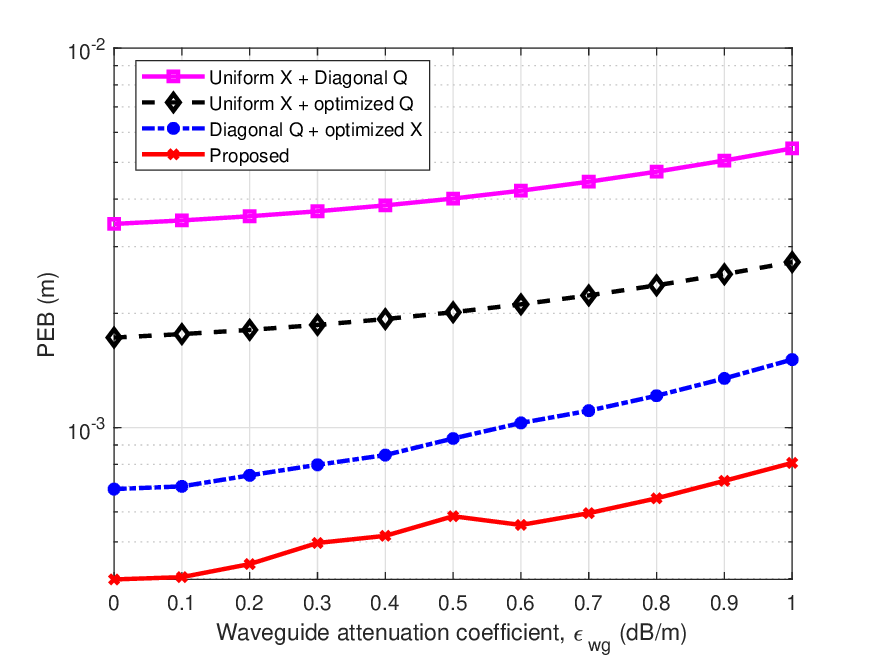}}
\caption{PEB versus the waveguide attenuation coefficient, with $D_x=50$ m, $P=30$ dBm, and $\bar R=1$ bit/s/Hz.}
\label{fig:attenuation}
\end{figure}

\begin{figure}[t]
\centering
{\includegraphics[width=0.5\textwidth]{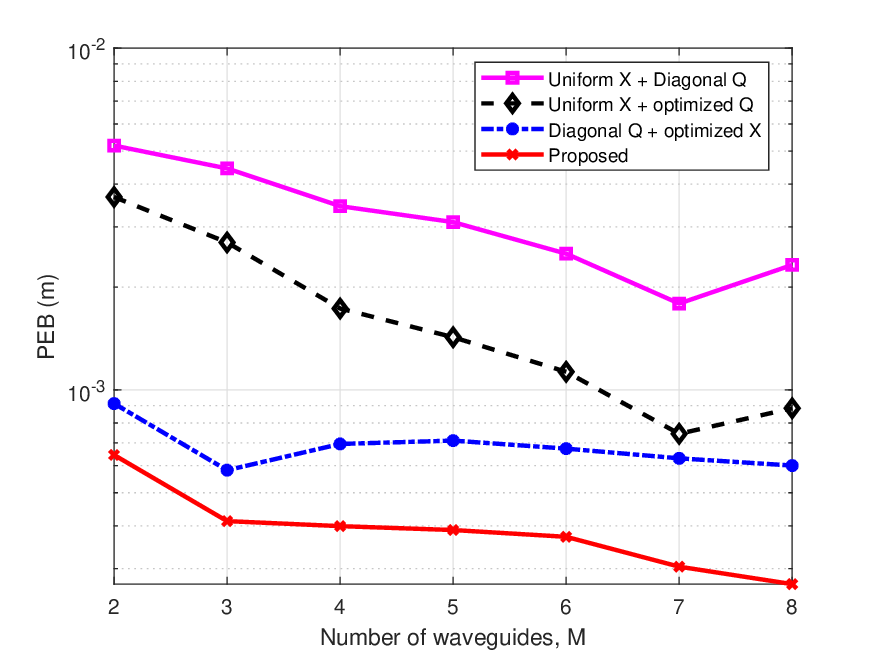}}
\caption{PEB versus the number of waveguides, with $N=6$, $D_y=2$ m, $P=30$ dBm, and $\bar R=1$ bit/s/Hz.}
\label{fig:number}
\end{figure}

\begin{figure}[t]
\centering
{\includegraphics[width=0.5\textwidth]{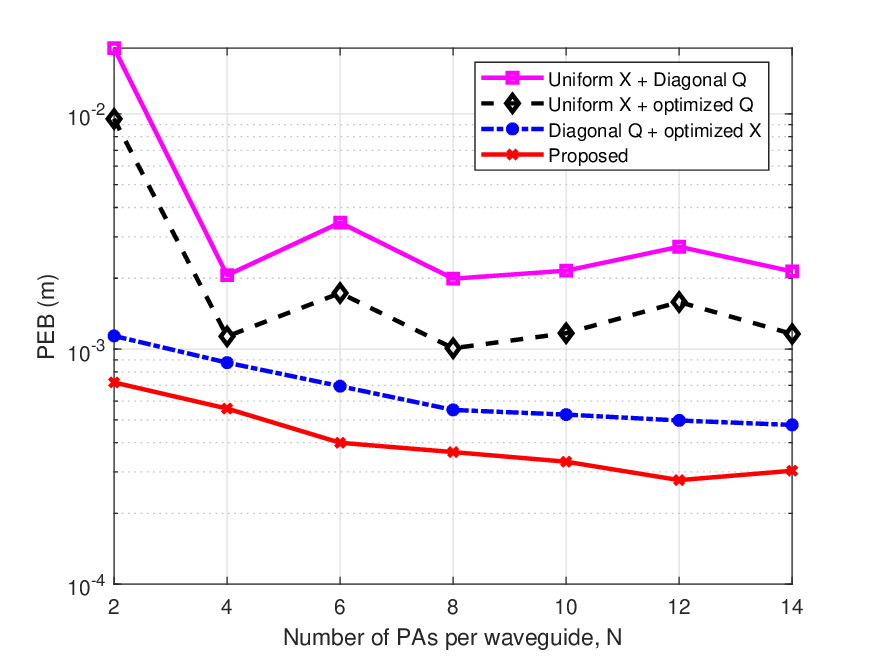}}
\caption{PEB versus the number of PAs per waveguide, with $M=4$, $D_x=50$ m, $P=30$ dBm, and $\bar R=1$ bit/s/Hz.}
\label{fig:perwaveguide}
\end{figure}

\section{Numerical Results}
\label{sec:numerical_results}

In this section, we evaluate the proposed joint transmit-covariance and PA-position design for a single near-field sensing target. Unless otherwise specified, the BS employs $M=4$ dielectric waveguides, each containing $N=6$ PAs, and serves $K=3$ multicast communication users. A co-located receive array with $N_{\mathrm r}=4$ antennas is used for monostatic sensing. The carrier frequency is $f_c=28$~GHz, the receive-antenna spacing is $\lambda/2$, the waveguide and receive-array heights are $h=h_r=3$~m, and the transverse waveguide aperture is $D_y=2$~m. The minimum PA spacing is $\Delta_{\min}=\lambda/2$. The communication and sensing noise powers are both set to $-94$~dBm, and $T_s=100$ sensing snapshots are used. A fixed realization of the user and target locations is employed for all schemes at each operating point so that the observed differences are caused by the resource and geometry designs rather than by different spatial realizations.

Because the steering vectors in \eqref{eq:at_entry} and \eqref{eq:ar_entry} already include the transmit- and receive-path distance factors, the target coefficient is set according to
\begin{IEEEeqnarray}{rCl}
\beta=|\beta|e^{j\phi_\beta}, ~ |\beta| &=& \sqrt{\frac{G_tG_r\lambda^2\sigma_{\rm rcs}}{(4\pi)^3L_s}}
=2.4052\times10^{-4},
\label{eq:beta_simulation}
\end{IEEEeqnarray}
with $G_t=G_r=L_s=\sigma_{\rm rcs}=1$ and $\phi_\beta=0.2$~rad. The phase $\phi_\beta=0.2$~rad is chosen as an arbitrary representative nonzero scattering phase. For the considered single-target model with an unknown complex reflection coefficient, the position CRB depends on $|\beta|$, rather than its particular phase.
Unless stated otherwise, the transmit-power budget is $P=30$~dBm, the multicast-rate threshold is $\bar R=1$~bit/s/Hz, the waveguide length is $D_x=50$~m, and the attenuation coefficient is $\varepsilon=0.01$~dB/m.

For the single-target experiments, the full FIM associated with $\boldsymbol{\xi}=[r,\theta,\Re\{\beta\},\Im\{\beta\}]^{\mathsf T}$ is retained so that the unknown reflection coefficient is treated as a nuisance parameter. Let $\mathbf C_{\boldsymbol\xi}=\mathbf J_{\boldsymbol\xi}^{-1}$. We evaluate the Cartesian position-oriented CRB and its corresponding position error bound (PEB) as
\begin{IEEEeqnarray}{rCl}
\operatorname{CRB}_{\rm pos}(\mathbf X,\mathbf Q)
&=& [\mathbf C_{\boldsymbol\xi}]_{1,1}
+r^2[\mathbf C_{\boldsymbol\xi}]_{2,2},
\label{eq:position_crb_simulation}\\
\operatorname{PEB}(\mathbf X,\mathbf Q)
&=& \sqrt{\operatorname{CRB}_{\rm pos}(\mathbf X,\mathbf Q)}.
\label{eq:peb_simulation}
\end{IEEEeqnarray}
The optimization algorithms use $\operatorname{CRB}_{\rm pos}$ in $\mathrm{m}^2$, whereas the performance figures report the PEB in meters for direct physical interpretation. The following four schemes are compared:
\begin{itemize}
    \item \textbf{Uniform $X$ + Diagonal $Q$}: the PAs are uniformly deployed and $\mathbf Q=(P/M)\mathbf I_M$;
    \item \textbf{Uniform $X$ + Optimized $Q$}: the PA positions are fixed uniformly and the full covariance matrix is optimized;
    \item \textbf{Diagonal $Q$ + Optimized $X$}: $\mathbf Q=(P/M)\mathbf I_M$ and the PA positions are optimized; and
    \item \textbf{Proposed}: the PA-position matrix and full covariance matrix are jointly optimized.
\end{itemize}
For the proposed design, the feasible optimized-$X$/diagonal-$Q$ solution is used as a geometry-informed initialization and the best feasible point is retained. This avoids degradation caused solely by an unfavorable initialization of the non-convex alternating procedure. Infeasible operating points are omitted from the figures.

Fig.~\ref{fig:convergence} illustrates the convergence behavior of the proposed AO algorithm for $(P=30)$ dBm. Iteration (0) corresponds to the geometry-informed feasible initialization obtained with optimized PA positions and diagonal covariance. Starting from a position CRB of approximately ($4.83\times10^{-7},\mathrm{m}^2$), the first complete AO iteration, consisting of a covariance update followed by the waveguide-wise PA-position updates, reduces the CRB to approximately ($1.59\times10^{-7},\mathrm{m}^2$). The objective then remains essentially unchanged over the subsequent iterations until the convergence criterion is satisfied. The non-increasing evolution is consistent with the acceptance safeguards used in the proposed algorithm and demonstrates its rapid convergence for the considered realization.


Fig.~\ref{fig:Power} shows the PEB versus the maximum transmit power. For fixed PA geometry, the FIM in \eqref{eq:FIM_reduced} is linear in $\mathbf{Q}$. Therefore, when the optimized covariance approximately scales with the available power and the active constraint structure remains unchanged, the FIM scales approximately linearly with $P$, resulting in an approximate $1/P$ scaling of the position CRB and a $1/\sqrt{P}$ scaling of the PEB. Since the PA geometry is jointly reoptimized and the multicast constraints may become active at different power levels, these scaling laws should be interpreted as approximate trends.

Fig.~\ref{fig:Rate} shows the PEB versus the multicast-rate threshold, where $P=10$~dBm is adopted to make the communication constraint active within the considered range. At low rate thresholds, the constraint is inactive or weakly active, and the PEB remains nearly constant. As $\bar{R}$ increases, more spatial and covariance degrees of freedom are devoted to satisfying the multicast users,
leaving less flexibility for sensing and increasing the PEB. This trend is also consistent with the communication--sensing phase conflict discussed in Section~III-E: communication-favorable PA
positions that enhance coherent signal accumulation in~\eqref{eq:comm_power_wg} need not preserve the phase diversity that improves the conditioning of
the sensing FIM in~\eqref{eq:Psi_mm_expanded}. This phase conflict can therefore partially contribute to the observed PEB increase, although Fig.~\ref{fig:Rate} does not isolate it from the general reduction in sensing-design freedom.
The fixed uniform-$\mathbf{X}$/diagonal-$\mathbf{Q}$ benchmark remains unchanged while feasible because it has no optimization variables. Once its maximum achievable multicast rate is exceeded, no feasible solution exists and hence no PEB value is reported.
The proposed joint design achieves the lowest PEB over its feasible range. The small fluctuations in the optimized-position curves are caused by the nonconvex position-search procedure.


Fig.~\ref{fig:length} illustrates the effect of the waveguide length $D_x$ for $\varepsilon=0.01$~dB/m.
Increasing $D_x$ enlarges the feasible PA-placement region. Since the feasible set for a shorter waveguide is contained in that of a longer waveguide, the globally optimal joint-design PEB cannot increase with $D_x$; the optimizer may always retain the previous PA locations. The gradual saturation indicates that the additional available length provides diminishing geometric benefit and is consistent with the optimizer using only an effective subregion of the waveguide to avoid unnecessary propagation loss. In contrast, the uniform-placement benchmarks redistribute the PAs over the entire available length. Their behavior reflects both increasing propagation loss and changes in the coherent phase relationships, and is therefore not necessarily monotonic. The small increases observed for the optimized schemes are attributable to the local nature of the geometry search rather than to enlargement of the global feasible set.


Fig.~\ref{fig:attenuation} plots the PEB versus the waveguide attenuation coefficient. The PEB exhibits an overall increasing trend because a larger $\varepsilon$ reduces the power reaching PAs located farther from the feed and weakens their contributions to the sensing FIM. The proposed design remains the best over the entire attenuation range by moving the PAs toward more favorable locations and adapting the full covariance matrix. Nevertheless, even the proposed PEB approximately doubles as $\varepsilon$ increases from $0$ to $1$~dB/m, confirming that waveguide loss cannot be ignored for electrically long PASS deployments. 
This sweep isolates the loss component identified in Section~III-C. As $\epsilon$ increases, the loss factors $\rho_m$ decrease, weakening both the per-waveguide and cross-waveguide FIM terms in~\eqref{eq:Psi_mm_expanded} and~\eqref{eq:Psi_mmp_expanded}. PA-position and covariance optimization can mitigate this loss by avoiding unfavorable distant locations and reallocating the feed-domain covariance, but they cannot eliminate the attenuation penalty. Together, Figs.~4 and~5 separately illustrate the aperture benefit and the guided-wave loss penalty underlying the loss--aperture mechanism.

Fig.~\ref{fig:number} shows the PEB versus the number of waveguides $M$ with $N=6$ PAs on each waveguide. Increasing $M$ adds transverse spatial samples, enlarges the dimension of the feed-domain covariance matrix, and increases the total number of PAs from $MN=12$ to $48$. The proposed PEB decreases from approximately $0.645$~mm at $M=2$ to $0.270$~mm at $M=8$. The gain gradually diminishes because both the total power and the transverse aperture $D_y$ remain fixed. The benchmark curves can exhibit mild non-monotonicity because changing $M$ also changes the uniform waveguide spacing and, for diagonal $\mathbf Q$, the power assigned to each waveguide. The persistent advantage of the proposed method verifies the benefit of jointly exploiting transverse geometry and the off-diagonal covariance entries.

Fig.~\ref{fig:perwaveguide} depicts the PEB versus the number of PAs per waveguide $N$. Increasing $N$ generally improves the optimized designs because more PAs create additional guided-wave and free-space phase diversity. The proposed design reduces the PEB from roughly $0.7$--$0.8$~mm at $N=2$ to approximately $0.3$~mm for the largest tested values of $N$. The improvement becomes progressively smaller because the feed power per waveguide is fixed and the additional PAs provide increasingly correlated information. The uniform-position benchmarks exhibit stronger fluctuations since inserting uniformly spaced PAs changes the entire coherent superposition and may create unfavorable phase combinations. The small variation at the largest $N$ values is also consistent with the local nature of the position optimization.

\vspace{-2pt}

\section{Conclusion}
This paper investigated a near-field multicast ISAC system enabled by a lossy multi-waveguide PASS. A unified model was developed by incorporating PA geometry, waveguide attenuation, equal-radiated-power operation, multicast downlink transmission, and monostatic sensing. To characterize the sensing performance, the average FIM and the corresponding CRB were derived for estimating the target locations and reflection coefficients. By exploiting the block-diagonal structure of the PA transfer matrix, a compact projected-Jacobian representation was obtained, which reveals how the PA positions and feed-domain transmit covariance jointly affect the sensing accuracy.
Based on this representation, we showed that the diagonal entries of the covariance matrix determine the per-waveguide sensing contributions, while the off-diagonal entries control the cross-waveguide correlation terms. We also identified the loss-aperture tradeoff caused by waveguide attenuation and the phase conflict between multicast communication and near-field sensing. To balance these effects, a joint PA-position and transmit-covariance optimization problem was formulated to minimize the CRB subject to multicast-rate, power, and PA-deployment constraints. An alternating algorithm was then developed, where the covariance subproblem is solved as an SDP and the PA-position subproblem is handled by waveguide-wise block coordinate descent. 
Numerical results demonstrated that the proposed joint design consistently outperforms benchmark schemes with fixed PA positions or diagonal covariance matrices. The results confirmed that PA-position optimization provides substantial sensing gains, while full covariance optimization further improves the CRB by exploiting cross-waveguide correlation. Future work may consider robust designs under imperfect location information, discrete PA activation, hardware impairments, and geometry-dependent radiation patterns.

\appendices

\section{Derivation of the Jacobian Matrix $\mathbf{D}_{\boldsymbol{\xi}}(\mathbf{X})$}
\label{app:jacobian}

This appendix derives in full the Jacobian matrix $\mathbf{D}_{\boldsymbol{\xi}}(\mathbf{X}) = \partial\bar{\boldsymbol{\mu}}(\boldsymbol{\xi},\mathbf{X})/\partial\boldsymbol{\xi}^{\mathsf{T}} \in \mathbb{C}^{N_{\mathrm{r}}N_{\mathrm{t}}\times 4L}$ defined in~\eqref{eq:Dxi_def}. Differentiating $\bar{\boldsymbol{\mu}}(\boldsymbol{\xi},\mathbf{X}) = \sum_{\ell=1}^{L}\beta_\ell\,\mathbf{a}_{\mathrm{t}}(\mathbf{p}_\ell,\mathbf{X}) \otimes \mathbf{a}_{\mathrm{r}}(\mathbf{p}_\ell)$ from~\eqref{eq:mubar_sum} with respect to each scalar in $\boldsymbol{\xi}$ yields, where $\otimes$ denotes the Kronecker product:
\begin{IEEEeqnarray}{rCl}
	\mathbf{d}_{r_\ell} = \frac{\partial \bar{\boldsymbol{\mu}}}{\partial r_\ell} &=& \beta_\ell\!\left(
	\frac{\partial \mathbf{a}_{\mathrm{t}}}{\partial r_\ell}
	\otimes \mathbf{a}_{\mathrm{r}}
	+ \mathbf{a}_{\mathrm{t}}
	\otimes \frac{\partial \mathbf{a}_{\mathrm{r}}}{\partial r_\ell}
	\right), \label{eq:dmu_dr} \\
	\mathbf{d}_{\theta_\ell} = \frac{\partial \bar{\boldsymbol{\mu}}}{\partial {\theta_\ell}} &=& \beta_\ell\!\left(
	\frac{\partial \mathbf{a}_{\mathrm{t}}}{\partial \theta_\ell}
	\otimes \mathbf{a}_{\mathrm{r}}
	+ \mathbf{a}_{\mathrm{t}}
	\otimes \frac{\partial \mathbf{a}_{\mathrm{r}}}{\partial \theta_\ell}
	\right), \label{eq:dmu_dtheta} \\
	\frac{\partial \bar{\boldsymbol{\mu}}}{\partial \Re\{\beta_\ell\}}
	&=& \mathbf{a}_{\mathrm{t}}(\mathbf{p}_\ell,\mathbf{X})
	\otimes \mathbf{a}_{\mathrm{r}}(\mathbf{p}_\ell),
	\label{eq:dmu_drebeta} \\
	\frac{\partial \bar{\boldsymbol{\mu}}}{\partial \Im\{\beta_\ell\}}
	&=& j\,\mathbf{a}_{\mathrm{t}}(\mathbf{p}_\ell,\mathbf{X})
	\otimes \mathbf{a}_{\mathrm{r}}(\mathbf{p}_\ell).
	\label{eq:dmu_dimbeta}
\end{IEEEeqnarray}
Please note that the derivatives with respect to $\Re\{\beta_\ell\}$ and $\Im\{\beta_\ell\}$ are immediate since $\bar{\boldsymbol{\mu}}$ is linear in $\beta_\ell$. 
The position derivatives in~\eqref{eq:dmu_dr} and \eqref{eq:dmu_dtheta} require the steering-vector entry derivatives $\partial[\mathbf{a}_{\mathrm{t}}]_{(m-1)N+n}/\partial\tilde{\xi}$ and $\partial[\mathbf{a}_{\mathrm{r}}]_q/\partial\tilde{\xi}$. Given the transmit and receive propagation distances appearing in~\eqref{eq:nearfield_dists_tx} and \eqref{eq:nearfield_dists_rx} can be re-written as
\begin{IEEEeqnarray}{rCl}
	d^{(\mathrm{t})}_{m,n,\ell}
	&=& \sqrt{\Delta x_{m,n,\ell}^2
		+ \Delta y_{m,\ell}^2 + h^2},
	\label{eq:dt_3d} \\
	d^{(\mathrm{r})}_{q,\ell}
	&=& \sqrt{\Delta x_{\mathrm{r},q,\ell}^2
		+ \Delta y_{\mathrm{r},q,\ell}^2 + h_r^2},
	\label{eq:dr_3d}
\end{IEEEeqnarray}
where $\Delta x_{m,n,\ell} = x_{m,n} - r_\ell\cos\theta_\ell$, $\Delta y_{m,\ell} = y_m - r_\ell\sin\theta_\ell$, $\Delta x_{\mathrm{r},q,\ell} = x_{\mathrm{r},q} - r_\ell\cos\theta_\ell$, and $\Delta y_{\mathrm{r},q,\ell} = y_{\mathrm{r},q} - r_\ell\sin\theta_\ell$. And since each steering-vector entry has the form $[\mathbf{a}]_i = e^{-j\kappa d_i}/d_i$, its derivative with respect to a scalar $\tilde{\xi} \in \{r_\ell, \theta_\ell\}$ follows by the chain rule as
\begin{IEEEeqnarray}{rCl}
	\frac{\partial [\mathbf{a}_{\mathrm{t}}]_{(m-1)N+n}}{\partial \tilde{\xi}}
	&=& - [\mathbf{a}_{\mathrm{t}}]_{(m-1)N+n}
	\!\left(\frac{1}{d^{(\mathrm{t})}_{m,n,\ell}}+j\kappa\right)
	\frac{\partial d^{(\mathrm{t})}_{m,n,\ell}}{\partial \tilde{\xi}},
	\label{eq:d_at_dxi} \nonumber \\ && \IEEEeqnarraynumspace \\
	\frac{\partial [\mathbf{a}_{\mathrm{r}}]_q}{\partial \tilde{\xi}}
	&=& -[\mathbf{a}_{\mathrm{r}}]_q
	\left(\frac{1}{d^{(\mathrm{r})}_{q,\ell}}+j\kappa\right)
	\frac{\partial d^{(\mathrm{r})}_{q,\ell}}{\partial \tilde{\xi}}.
	\label{eq:d_ar_dxi}
\end{IEEEeqnarray}
The required first-order distance derivatives are obtained by differentiating~\eqref{eq:dt_3d} and \eqref{eq:dr_3d}. Since $h$ and $h_r$ are independent of $(r_\ell, \theta_\ell, x_{m,n})$, they contribute only through the denominators $d^{(\mathrm{t})}_{m,n,\ell}$ and $d^{(\mathrm{r})}_{q,\ell}$, respectively. This gives
\begin{IEEEeqnarray}{rCl}
	\frac{\partial d^{(\mathrm{t})}_{m,n,\ell}}{\partial r_\ell}
	&=& \frac{
		-\Delta x_{m,n,\ell}\cos\theta_\ell
		- \Delta y_{m,\ell}\sin\theta_\ell
	}{d^{(\mathrm{t})}_{m,n,\ell}},
	\label{eq:ddt_dr} \\
	\frac{\partial d^{(\mathrm{t})}_{m,n,\ell}}{\partial\theta_\ell}
	&=& \frac{
		r_\ell\!\left(
		\Delta x_{m,n,\ell}\sin\theta_\ell
		- \Delta y_{m,\ell}\cos\theta_\ell
		\right)
	}{d^{(\mathrm{t})}_{m,n,\ell}},
	\label{eq:ddt_dtheta} \\
	\frac{\partial d^{(\mathrm{r})}_{q,\ell}}{\partial r_\ell}
	&=& \frac{
		-\Delta x_{\mathrm{r},q,\ell}\cos\theta_\ell
		- \Delta y_{\mathrm{r},q,\ell}\sin\theta_\ell
	}{d^{(\mathrm{r})}_{q,\ell}},
	\label{eq:ddr_dr} \\
	\frac{\partial d^{(\mathrm{r})}_{q,\ell}}{\partial\theta_\ell}
	&=& \frac{
		r_\ell\!\left(
		\Delta x_{\mathrm{r},q,\ell}\sin\theta_\ell
		- \Delta y_{\mathrm{r},q,\ell}\cos\theta_\ell
		\right)
	}{d^{(\mathrm{r})}_{q,\ell}}.
	\label{eq:ddr_dtheta}
\end{IEEEeqnarray}

Substituting~\eqref{eq:ddt_dr}--\eqref{eq:ddr_dtheta} into~\eqref{eq:d_at_dxi} and \eqref{eq:d_ar_dxi} gives the first-order steering-vector entry derivatives, which upon insertion into~\eqref{eq:dmu_dr} and \eqref{eq:dmu_dtheta} yield the position-related columns $\mathbf{d}_{r_\ell}$ and $\mathbf{d}_{\theta_\ell}$ of $\mathbf{D}_{\boldsymbol{\xi}}(\mathbf{X})$. Additionally, stacking~\eqref{eq:dmu_drebeta} and~\eqref{eq:dmu_dimbeta} across all $L$ targets and recalling that the $\ell$-th column of the Khatri--Rao product $\mathbf{A}_{\mathrm{t}}\diamond\mathbf{A}_{\mathrm{r}}$ is by definition $\mathbf{a}_{\mathrm{t}}(\mathbf{p}_\ell,\mathbf{X})\otimes\mathbf{a}_{\mathrm{r}}(\mathbf{p}_\ell)$ directly yields the last two column groups of~\eqref{eq:Dxi_closed}, namely $\mathbf{A}_{\mathrm{t}}\diamond\mathbf{A}_{\mathrm{r}}$ and $j\,\mathbf{A}_{\mathrm{t}}\diamond\mathbf{A}_{\mathrm{r}}$. This fully specifies the Jacobian matrix $\mathbf{D}_{\boldsymbol{\xi}}(\mathbf{X})$ in closed form as stated in~\eqref{eq:Dxi_closed}.

\bibliographystyle{IEEEtran}
\bibliography{biblio}

\end{document}